\documentclass[prb,aps,preprint,amssymb,superscriptaddress]{revtex4-1}
\usepackage{graphicx}
\usepackage{dcolumn}
\usepackage{bm}
\usepackage[usenames,dvipsnames]{color}
\usepackage{amsmath}
\newcommand{\tg}{}
\newcommand{\tb}{}
\newcommand{\tr}{}

\newcommand{\bk}{\mathbf{k}}
\newcommand{\ua}{\uparrow}
\newcommand{\da}{\downarrow}
\newcommand{\eq}{\begin{equation}}
\newcommand{\eqx}{\end{equation}}
\newcommand{\eqn}{\begin{eqnarray}}
\newcommand{\eqnx}{\end{eqnarray}}
\newcommand{\s}{\sigma}
\newcommand{\veck}{{\bf k}}
\newcommand{\veci}{{\bf i}}
\newcommand{\vecj}{{\bf j}}

\newcommand{\vecm}{{\bf m}}
\newcommand{\vecl}{{\bf l}}
\newcommand{\ra}{\rangle}
\newcommand{\la}{\langle}
\newcommand{\lam}{\lambda}
\newcommand{\ov}{\overline}
\newcommand{\e}{\emptyset}
\newcommand{\dg}{\dagger}
\newcommand{\dsbeg}{\sum_{\vecl_1,\ldots, \vecl_k} \bigl\langle}
\newcommand{\dsend}{\hat{d}^{\rm HF}_{\vecl_1,\ldots,\vecl_k} \bigr\rangle^{\rm c}_{0}}
\newcommand{\Ps}{|\Psi_0\rangle}

\begin{document}

\title{High temperature superconductivity in the two-dimensional $t$-$J$ model: Gutzwiller wave function solution}

\author{Jan Kaczmarczyk}
\email{jan.kaczmarczyk@uj.edu.pl}
\affiliation{Instytut Fizyki im. Mariana Smoluchowskiego, Uniwersytet Jagiello\'{n}ski, Reymonta 4, 30-059 Krak\'ow, Poland}

\author{J\"org B\"unemann}
\email{buenemann@googlemail.com}
\affiliation{Institut f\"ur Theoretische Physik, Leibniz Universit\"at, D-30167 Hannover, Germany}

\author{J\'{o}zef Spa\l ek}
\email{ufspalek@if.uj.edu.pl}
\affiliation{Instytut Fizyki im. Mariana Smoluchowskiego, Uniwersytet Jagiello\'{n}ski, Reymonta 4, 30-059 Krak\'ow, Poland}

\date{\today}

\begin{abstract}%
A systematic diagrammatic expansion for Gutzwiller-wave functions (DE-GWF) \tg{proposed very recently} is used for the description of superconducting (SC) ground state in the two-dimensional square-lattice $t$-$J$ model with the hopping electron amplitudes~$t$ (and~$t'$) between nearest (and next-nearest) neighbors. \tg{On the example of the SC state analysis we provide a detailed comparison of the method results with other approaches. Namely: (i) the truncated DE-GWF method reproduces the variational Monte Carlo (VMC) results; (ii) in the lowest (zeroth) order of the expansion the method can reproduce the analytical results of the standard Gutzwiller approximation (GA), as well as of the recently proposed ``grand-canonical Gutzwiller approximation'' (GCGA).} We obtain important features of the SC state. First, the SC gap at the Fermi surface resembles a $d_{x^2-y^2}$-wave only \tg{for optimally- and overdoped system, being diminished in the antinodal regions for the underdoped case in a qualitative agreement with experiment.} Corrections to the gap structure are shown to arise from the longer range of the real-space pairing. Second, the nodal Fermi velocity is almost constant as a function of doping and agrees semi-quantitatively with experimental results. Third, we compare the doping dependence of the gap magnitude with experimental data. Fourth, we analyze the $\bk$-space properties of the model: \tg{Fermi surface topology and effective dispersion. The DE-GWF method opens up new perspectives for studying strongly-correlated systems, as: (i) it works in the thermodynamic limit, (ii) is comparable in accuracy to VMC, and (iii) has numerical complexity comparable to GA (i.e., it provides the results much faster than the VMC approach). }
 \end{abstract}



\maketitle

\section{Introduction}

The Hubbard and the $t$-$J$ models of strongly correlated fermions play an eminent role in rationalizing the principal properties of high temperature superconductors (for recent reviews see \cite{Rev_High_Tc,Ogata,RevModPhys.84.1383,EdeggerRev,Science.317.1704}). The relative role of the particles' correlated motion and the binding provided by the kinetic exchange interaction can be clearly visualized in the effective $t$-$J$ model, where the effective hopping energy $\sim |t|\delta \sim 0.35 \, \rm{eV}$ ($\delta \equiv 1-2n$ is the hole doping) is comparable or even lower than the kinetic exchange integral $J \approx 0.12 \, \rm{eV}$. Simply put, the hopping electron drags behind its exchange-coupled nearest neighbor (n.n.) via empty sites and thus preserves the locally bound configuration in such correlated motion throughout the lattice \cite{JSDGJ}. In effect, this real-space pairing picture is complementary to the more standard virtual boson (paramagnon) exchange mechanism which involves, explicitly or implicitly, a quasiparticle picture and concomitant with it reciprocal-space language \cite{PhysRevLett.96.047005,*PhysRevB.74.094513,PhysRevLett.67.3448,PhysRevB.80.205109,Hanke1}. Unfortunately, no single unifying approach, if possible at all, exists in the literature which would unify the Eliashberg-type and the real-space approaches, out of which a Cooper-pair condensate would emerge as a universal state for arbitrary ratio of the band energy $\sim W$ to the Coulomb repulsion $U$. The reason for this exclusive character of the approaches is ascribed to the presence of the Mott-Hubbard phase transition that takes place for $W/U \approx 1$ (appearing for the half-filled band case) which also delineates the strong-correlation limit for a doped-Mott metallic state, for $W$ substantially smaller than $U$. This is the regime, where the $t$-$J$ model is assumed to be valid, even in the presence of partially-filled oxygen $2p$ states \cite{Hanke1,AndersonBook,PhysRevB.37.3759,*PhysRevB.41.7243,Spalek1,*Spalek2}. The validity of this type of physics is assumed throughout the present paper and a quantitative analysis of selected experimental properties, as well as a comparison with variational Monte-Carlo (VMC) results, is undertaken.

One of the approaches designed to interpolate between the $W/U \gg 1$ and $W/U \lesssim 1$ limits is the Gutzwiller wave function (GWF) approach \cite{Gutzwiller,*PhysRev.137.A1726}. Unfortunately, the method does not allow for an extrapolation to the $W/U \ll 1$ limit, at least in the simpler Gutzwiller approximation (GA) \cite{Zhang}. Therefore, different forms of the GA-like approaches, appropriate for the $t$-$J$ model, have been invented under the name of the Renormalized Mean Field Theory (RMFT) \cite{Zhang,Ogata2,Wang,Fukushima,Fukushima2,Jedrak2,EdeggerRev}. The last approach based on the $t$-$J$ model provides a rationalization of the principal characteristics of high temperature superconductors, including selected properties in a semiquantitative manner, particularly when the so-called statistically consistent Gutzwiller approach (SGA) \cite{Jedrak1,Jedrak2,PhysRevB.84.125140,Olga,Zegrodnik,Abram} is incorporated into RMFT. However, one should also mention that neither GA nor SGA provide a stable superconducting state in the Hubbard model.

Under these circumstances, we have undertaken a project involving a full GWF solution via a systematic Diagrammatic Expansion of the GWF (DE-GWF), which becomes applicable to two- and higher-dimensional systems, for both normal\cite{Buenemann} and superconducting\cite{PhysRevB.88.115127} states. Previously, this solution has been achieved in one-spatial dimension in an iterative manner \cite{Metzner,Kurzyk}. Obviously, the DE-GWF should reduce to SGA in the limit of infinite dimensions. In our preceding paper\cite{PhysRevB.88.115127} we have presented the first results for the Hubbard model. Here, a detailed analysis \tb{is provided for the $t$-$J$ model}, together with a comparison to experiment, as well as to the VMC and GA results. The limitations of the present approach are also discussed, particularly the inability to describe the pseudogap appearance.

The structure of the paper is as follows. In Sec. II we present the DE-GWF method (cf. also Appendices A and B). In Secs. III and IV (cf. also Appendices C, D, and E) we provide details of the numerical analysis and discuss physical results, respectively. In the latter section, we also compare our results with experiment and relate them to VMC and GA results. Finally, in Sec. V we draw conclusions and overview our approach.

\section{The method}

\subsection{$t$-$J$ Model}

We start with the $t$-$J$ model Hamiltonian
\footnote{for original derivation of the $t$-$J$ model from the Hubbard model see: K. A. Chao, J. Spa{\l}ek, and A. M. Ole\'{s}, J. Phys. C \textbf{10}, L271 (1977). For didactical exposition see: J. Spa{\l}ek, Acta Phys. Polon. A \textbf{121}, 764 (2012).} \addtocounter{footnote}{-1} 
on a two-dimensional, infinite square lattice
\eqn
\hat{H}&=&\hat{H}_0 + \hat{H}_{ex} \,, \label{eq:H} \\
\hat{H}_0&=&\sum_{\veci,\vecj,\sigma}t_{\veci\vecj}\hat{c}_{\veci,\sigma}^{\dagger} (1 - \hat{n}_{\veci \overline{\s}}) \hat{c}_{\vecj,\sigma}^{\phantom{\dagger}} (1 - \hat{n}_{\vecj \overline{\s}}), \\
\hat{H}_{ex} &=& J \sum_{\la \veci, \vecj \ra} \left( \mathbf{\hat{S}_\veci}\mathbf{\hat{S}_\vecj} - c \frac{1}{4} \hat{\nu}_\veci \hat{\nu}_\vecj \right),
\eqnx
where $\hat{\nu}_\veci = \hat{\nu}_{\veci\ua} + \hat{\nu}_{\veci\da}$ with $\hat{\nu}_{\veci\s} \equiv \hat{n}_{\veci \s}(1-\hat{n}_{\veci \ov{\s}})$, the first term is the kinetic energy part and the second expresses the kinetic exchange. The spin operator is defined as $\mathbf{\hat{S}_\veci} = \{\hat{S}_\veci^z, \hat{S}_\veci^{+}, \hat{S}_\veci^{-}\}$ and $\sum_{\la \veci, \vecj \ra}$ denotes summation over pairs of n.n. sites (bonds). The parameter $c$ is used to switch on ($c=1$) or off ($c=0$) the density-density interaction term reproducing the two forms of the model used in the literature. Unless stated otherwise, the system's spin-isotropy and the translational invariance are not assumed and the analytical results presented in this section are valid for phases with broken symmetries.
\tg{We study system properties in the thermodynamic limit, i.e., the system size $L$ is infinite.}
We also neglect the three-site terms\footnotemark.

\subsection{Trial wave function}

The principal task within a Gutzwiller-type\cite{Gutzwiller} of approach is the calculation of the expectation value of the starting Hamiltonian with respect to the trial wave function, which is defined as
\begin{equation}
|\Psi_{\rm G}\rangle = \hat{P} |\Psi_0\rangle \equiv \prod\nolimits_{\veci}\hat{P}_{\veci}|\Psi_0\rangle \; ,
\label{eq:1.2}
\end{equation}
where $|\Psi_0\rangle$ is a single-particle product state (Slater determinant) to be specified later. We define the local Gutzwiller correlator in the atomic basis of the form
\begin{equation}
\hat{P}_{\veci} \equiv
\sum_{\Gamma}\lambda_{\veci,\Gamma} |\Gamma \rangle_{\veci\,\veci}\! \langle
\Gamma | \; , \label{eq:proj1}
\end{equation}
with variational parameters $\lambda_{\veci, \Gamma} \in \left\{ \lambda_{\veci, \emptyset}, \lambda_{\veci, 1\ua}, \lambda_{\veci, 1\da}, \lambda_{\veci, d} \right\}$, which describe the occupation probabilities of the four possible local states $ \{ |\Gamma\rangle_{\veci} \} \equiv \left\{|\emptyset\rangle_{\veci}, |\uparrow\rangle_{\veci}, |\downarrow\rangle_{\veci}, |\uparrow\downarrow\rangle_{\veci}\right\}$.
A particularly useful choice of the parameters $\lambda_{\veci,\Gamma}$ is the one which obeys
\eq
\hat{P}^2_{\veci} \equiv 1 + x \hat{d}_{\veci}^{\rm HF} \; , \label{eq:proj2}
\eqx
where the Hartree--Fock operators are defined by $\hat{d}_{\veci}^{\rm HF} \equiv \hat{n}^{\rm HF}_{\veci\uparrow}\hat{n}^{\rm HF}_{\veci\downarrow}$ and $\hat{n}^{\rm HF}_{\veci\sigma} \equiv \hat{n}_{\veci,\sigma} - n_{\veci\s} $ with $n_{\veci\s} = \langle \Psi_0| \hat{n}_{\veci\sigma} |\Psi_0 \rangle$. This form of $\hat{P}^2_{\veci}$ decisively simplifies the calculations by eliminating the so-called `Hartree bubbles'~\cite{Gebhard,Buenemann}.

For the t-J model, we work with zero double-occupancy, which sets $\lambda_{\veci, d} = 0$ and eliminates~$x$ as a variational parameter from the solution procedure. Explicitly, from the conditions in Eqs. (\ref{eq:proj1}) and (\ref{eq:proj2}) we find $\lam_{\veci, d}^2 = 1 + x (1-n_{\veci\ua})(1-n_{\veci\da}) = 0$. Calculating $x$ and inserting to the expressions for $\lam_{\veci, 1\s}$ and $\lam_{\veci, \e}$ gives
\eqn
\lam_{\veci, 1\s} &=& \frac{1}{\sqrt{1-n_{\veci\ov{\s}}}}, \\
\lam_{\veci, \e} &=& \sqrt{\frac{1-n_\veci}{(1-n_{\veci\s})(1-n_{\veci\ov{\s}})}},
\eqnx
where $n_\veci = n_{\veci\ua} + n_{\veci\da}$.

\subsection{Diagrammatic sums}

Here we discuss the analytical procedure of calculating the expectation value
\eq
W \equiv \la \hat{H} \ra_{\rm G} \equiv \frac{\la \Psi_{\rm G} | \hat{H} |\Psi_{\rm G} \ra}{\la \Psi_{\rm G} | \Psi_{\rm G} \ra} \equiv \frac{ \la \Psi_{0} | \hat{P} \hat{H} \hat{P} |\Psi_{0} \ra }{ \la \Psi_{0} | \hat{P}^2 |\Psi_{0} \ra } \label{eq:W}
\eqx
in detail for the kinetic-energy term and we summarize the results for other terms. We start with expressions for the relevant expectation values of interest via the power series in $x$, i.e.,
{\arraycolsep=2pt\begin{eqnarray}
\langle\Psi_{\rm G}|\Psi_{\rm G} \rangle = \Bigl\langle \prod_{\vecl}\hat{P}^2_{\vecl}\Bigr\rangle_{0}
&=&\sum_{k=0}^{\infty}\frac{x^k}{k!}
\sideset{}{'}\sum_{\vecl_1,\ldots, \vecl_k}
\bigl\langle \hat{d}^{\rm HF}_{\vecl_1,\ldots ,\vecl_k}
\bigr\rangle_{0}
\label{eq:1.9}\, ,\\
\langle\Psi_{\rm G}|\hat{c}^{\dagger}_{\veci,\sigma}\hat{c}_{\vecj,\sigma}^{\phantom{\dagger}}
|\Psi_{\rm G} \rangle = \Bigl\langle
\widetilde{c}^{\dagger}_{\veci,\sigma}\widetilde{c}_{\vecj,\sigma}^{\phantom{\dagger}}
\prod_{\vecl(\neq \veci,\vecj)}\hat{P}^2_{\vecl}\Bigr\rangle_{0}
&=&
\sum_{k=0}^{\infty}\frac{x^k}{k!}
\sideset{}{'}\sum_{\vecl_1,\ldots, \vecl_k}
\bigl\langle
\widetilde{c}_{\veci,\sigma}^{\dagger}
\widetilde{c}_{\vecj,\sigma}^{\phantom{\dagger}}
\hat{d}^{\rm HF}_{\vecl_1,\ldots,\vecl_k}\bigr\rangle_{0} \, ,
\label{eq:1.9c}
\end{eqnarray}}
where $\langle (...) \rangle_0 \equiv \langle \Psi_0| (...) |\Psi_0\rangle$, $\widetilde{c}_{\veci,\sigma}^{(\dagger)}\equiv
\hat{P}_{\veci}\hat{c}_{\veci,\sigma}^{(\dagger)}\hat{P}_{\veci}$, and we have defined
$
\hat{d}^{\rm HF}_{\vecl_1,\ldots,\vecl_k}\equiv\hat{d}^{\rm HF}_{\vecl_1}\cdots
\hat{d}^{\rm HF}_{\vecl_k}$
with
$
\hat{d}^{\rm HF}_{\emptyset}\equiv 1
$,
whereas the primed sums have the summation restrictions $\vecl_p\neq \vecl_{p'}$, $ \vecl_p \neq  \veci,\vecj$ for all $p,p'$.

Expectation values can now be evaluated by means of the Wick's theorem~\cite{Fetter} and are carried out in real space. Then, in the resulting diagrammatic expansion, the $k$-th order terms of Eqs.~(\ref{eq:1.9})-(\ref{eq:1.9c}) correspond to diagrams with one (or two) external vertices on sites $\veci$ (or $\veci$ and $\vecj$) and $k$ internal vertices. These vertices are connected with lines \tg{(corresponding to contractions from Wick's theorem)}, which in the case of the superconducting state with intersite pairing are given by
\eq    \label{pl}
P_{\vecl,\vecl'} \equiv P^{\sigma}_{\vecl,\vecl'}  \equiv
\langle\hat{c}^{\dagger}_{\vecl,\sigma}\hat{c}_{\vecl',\sigma}^{\phantom{\dagger}}\rangle_0
-\delta_{\vecl,\vecl'}n\;, \qquad
S_{\vecl,\vecl'}  \equiv
\langle\hat{c}^{\dagger}_{\vecl,\ua}\hat{c}^{\dagger}_{\vecl',\da}\rangle_0,
\eqx
where $\bar{\uparrow}=\downarrow$, $\bar{\downarrow}=\uparrow$. At this point, the application of the linked-cluster theorem~\cite{Fetter} yields~\cite{Buenemann} the analytical result for the kinetic energy term
\eq
\langle \hat{H}_0 \rangle_{\rm G} = \sum_{\veci,\vecj,\s} t_{\veci,\vecj} \left(
q_{\veci \s} q_{\vecj \s}               T_{\veci, \vecj, \s}^{(1),(1)}
+ q_{\veci \s} \alpha_{\vecj \s}        T_{\veci, \vecj, \s}^{(1),(3)}
+ \alpha_{\veci \s} q_{\vecj \s}        T_{\veci, \vecj, \s}^{(3),(1)}
+ \alpha_{\veci \s} \alpha_{\vecj \s}   T_{\veci, \vecj, \s}^{(3),(3)} \right), \label{eq:AvH0}
\eqx
where
\eqn
q_{\veci \s} &\equiv& \lambda_{\veci, 1\s} \lambda_{\veci, \emptyset} (1 - n_\veci) = \sqrt{\frac{1-n_\veci}{1-n_{\veci, \s}}}, \\
\alpha_{\veci \s} &\equiv& - \lambda_{\veci, 1\s} \lambda_{\veci, \emptyset} = - \frac{1}{1-n_{\veci\ov{\s}}} \sqrt{\frac{1-n_\veci}{1-n_{\veci, \s}}} = - \frac{q_{\veci \s}}{1-n_{\veci \ov{\s}}}.
\eqnx
The diagrammatic sums appearing in Eq. (\ref{eq:AvH0}) are defined by
\begin{equation}
S=\sum_{k=0}^{\infty}\frac{x^k}{k!} S(k), \label{eq:DS1}
\end{equation}
where
\begin{equation}
S\in\left\{T_{\veci, \vecj, \s}^{(1),(1)}, T_{\veci, \vecj, \s}^{(1),(3)}, T_{\veci, \vecj, \s}^{(3),(1)}, T_{\veci, \vecj, \s}^{(3),(3)} \right\}
\end{equation}
and the $k$-th order sum contributions have the following forms
\eq
T_{\veci,\vecj,\s}^{(1)[(3)],(1)[(3)]}(k) \equiv \dsbeg   [\hat{n}^{\rm HF}_{\veci,\bar{\sigma}}]
\hat{c}^{\dagger}_{\veci,\sigma} [\hat{n}^{\rm HF}_{\vecj,\bar{\sigma}}] \hat{c}_{\vecj,\sigma}^{\phantom{\dagger}} \dsend
\eqx
where $\langle \dots \rangle_0^{ \rm c}$ indicates that only the connected diagrams are to be kept (see Appendix \ref{app:A0} for exemplary diagrams and their contributions to diagrammatic sums in the two lowest orders). The notation $(1)[(3)]$ means that for the index (3) also the term in square brackets needs to be taken into account, e.g. $T_{\veci,\vecj, \s}^{(1),(3)}(k) \equiv
\sum_{\vecl_1,\ldots,\vecl_k}
\bigl\langle
\hat{c}^{\dagger}_{\veci,\sigma}
\hat{n}^{\rm HF}_{\vecj,\bar{\sigma}}
\hat{c}_{\vecj,\sigma}^{\phantom{\dagger}}\hat{d}^{\rm HF}_{\vecl_1,\ldots,\vecl_k}
\rangle^{\rm c}_{0}$. \tr{In the following expressions we will drop the brackets in the upper indices of diagrammatic sums for the sake of brevity.}

The exchange term can be rewritten in the form
\eq
J \sum_{\la \veci, \vecj \ra} \left( \mathbf{\hat{S}_\veci}\mathbf{\hat{S}_\vecj} - \frac{1}{4} \hat{\nu}_\veci \hat{\nu}_\vecj \right) =
J \sum_{\la \veci, \vecj \ra} \left(
\frac{\hat{S}_\veci^{+}\hat{S}_\vecj^{-} + \hat{S}_\veci^{-}\hat{S}_\vecj^{+}}{2}
+ \hat{S}_\veci^{z}\hat{S}_\vecj^{z}
- \frac{1}{4} \hat{\nu}_\veci \hat{\nu}_\vecj \right),
\eqx
\tg{where the spin-component operators are given by $\{\hat{S}_\veci^{+}, \hat{S}_\veci^{-}, \hat{S}_\veci^{z}\} = \{\hat{c}^\dg_{\veci\ua} \hat{c}_{\veci\da}, \hat{c}^\dg_{\veci\da} \hat{c}_{\veci\ua}, \frac{1}{2}(\hat{n}_{\veci\ua} - \hat{n}_{\veci\da}) \}$.}
%
The expectation values of the exchange term components can be expressed as
\eq
\frac{1}{2}\la \hat{S}_\veci^{+}\hat{S}_\vecj^{-} + \hat{S}_\veci^{-}\hat{S}_\vecj^{+} \ra_G = \left[(n_{{\veci\da}}-1) (n_{{\veci\ua}}-1) (n_{{\vecj\da}}-1)(n_{{\vecj\ua}}-1)\right]^{-1/2} \frac{S_{\veci\ua, \vecj\da}^{22}+S_{\veci\da, \vecj\ua}^{22}}{2}. \label{eq:S+S-}
\eqx
For the expressions of the other components see Appendix \ref{app:A}.

The diagrammatic sums appearing in the above expressions are defined by Eq. (\ref{eq:DS1}) with
\begin{equation}
S\in\left\{I_{\veci[\vecj]\s}^{2}, I_{\veci[\vecj]}^{4}, I_{\veci\s, \vecj\s'}^{22}, I_{\veci\s, \vecj}^{24}, I_{\veci, \vecj\s}^{42}, I_{\veci, \vecj}^{44}, S_{\veci\s, \vecj\ov{\s}}^{22}\right\}
\end{equation}
and the $k$-th order sum contributions of the following forms
\eqn
I_{\veci[\vecj]\s}^{2}(k)       &\equiv& \dsbeg   \hat{n}^{\rm HF}_{\veci[\vecj]\s}                         \dsend, \\
I_{\veci[\vecj]}^{4}(k)         &\equiv& \dsbeg   \hat{d}^{\rm HF}_{\veci[\vecj]}                           \dsend, \\
I_{\veci\s,\vecj\s'}^{22}(k)    &\equiv& \dsbeg   \hat{n}^{\rm HF}_{\veci\s} \hat{n}^{\rm HF}_{\vecj\s'}    \dsend, \\
I_{\veci\s, \vecj}^{24}(k)      &\equiv& \dsbeg   \hat{n}^{\rm HF}_{\veci\s} \hat{d}^{\rm HF}_{\vecj}       \dsend, \\
I_{\veci, \vecj\s}^{42}(k)      &\equiv& \dsbeg   \hat{d}^{\rm HF}_{\veci} \hat{n}^{\rm HF}_{\vecj\s}       \dsend, \\
I_{\veci, \vecj}^{44}(k)        &\equiv& \dsbeg   \hat{d}^{\rm HF}_{\veci} \hat{d}^{\rm HF}_{\vecj}         \dsend, \\
S_{\veci\s, \vecj\ov{\s}}^{22}(k)&\equiv& \dsbeg  \hat{c}^\dg_{\veci\s}\hat{c}_{\veci\ov{\s}} \hat{c}^\dg_{\vecj\ov{\s}}\hat{c}_{\vecj\s}  \dsend.
\eqnx
In what follows, we evaluate these diagrammatic sums in particular situations.

\subsection{Spin-isotropic case}

The above expressions simplify significantly when a system with translational invariance and spin isotropy is considered. Explicitly, they become
\eqn
\la \hat{c}^{\dagger}_{\veci,\sigma}\hat{c}_{\vecj,\sigma} \ra_G &=& q^2 T^{11} + 2 q \alpha T^{13} + \alpha^2 T^{33}, \label{eq:iso1} \\
\frac{1}{2}\la \hat{S}_\veci^{+}\hat{S}_\vecj^{-} + \hat{S}_\veci^{-}\hat{S}_\vecj^{+} \ra_G &=& g_s S^{22},  \label{eq:iso2} \\
\la \hat{S}_\veci^{z}\hat{S}_\vecj^{z} \ra_G &=& g_s \left(\frac{I^{22 \ua\ua} - I^{22 \ua \da}}{2}\right),  \label{eq:iso3} \\
\frac{1}{4} \la \hat{n}_\veci \hat{n}_\vecj \ra_G &=&
n^2
+ I^{22\ua\ua} \frac{\left(1-2 n\right){}^2}{2 \left(n-1\right){}^2}
+ I^{22\ua\da}\frac{ \left(1-2 n\right){}^2}{2 \left(n-1\right){}^2}
+ I^{44} \frac{\left(1-2   n\right){}^2}{\left(n-1\right){}^4} + \nonumber \\
&&
  I^{2} \frac{n (4n - 2)}{n-1}
+ I^{4} \frac{2 n \left(2n-1\right)}{\left(n-1\right){}^2}
+ I^{24} \frac{2 \left(1-2 n\right){}^2}{\left(n-1\right){}^3},\label{eq:iso4}
\eqnx
where $n = n_{\veci\s} = n_{\vecj\s}$, $g_s = \frac{1}{(1-n)^2}$, $q^2 \equiv g_t = (1-2n)/(1-n)$, $\alpha = -q / (1-n)$, and the diagrammatic sums have also been simplified with $I^{22\ua\ua} = I_{\veci\s,\vecj\s}^{22}$, $I^{22\ua\da} = I_{\veci\s,\vecj\ov{\s}}^{22}$, $I^{24} = I_{\veci\s, \vecj}^{24} = I_{\veci, \vecj\s}^{42}$, $I^{2} = I_{\veci[\vecj]\s}^{2}$, and $I^{4} = I_{\veci[\vecj]}^{4}$, $S^{22} = S_{\veci\s, \vecj\ov{\s}}^{22}$.

Note that the rotational-invariance requires $\la \hat{S}_\veci^{z}\hat{S}_\vecj^{z} \ra = \la \hat{S}_\veci^{x}\hat{S}_\vecj^{x} \ra = \la \hat{S}_\veci^{y}\hat{S}_\vecj^{y} \ra$, which leads to the condition for diagrammatic sums: $S^{22} = I^{22 \ua\ua} - I^{22 \ua \da}$. We have verified that this condition holds true in our calculations.

In general, this situation is applicable when no N\'{e}el-type antiferromagnetism occurs, as for the spin-singlet paired state the spin isotropy is preserved.

\subsection{Relation to other approaches} \label{sec:equiv}

When only the zeroth order of the diagrammatic expansion method is taken into account and under additional simplifications (see below), the analytical results are equivalent to those of the Gutzwiller approximation (GA) \cite{Zhang, EdeggerRev} and of the recently proposed grand-canonical Gutzwiller approximation (GCGA) \cite{Fukushima, Fukushima2, Jedrak1, Jedrak2, PhysRevB.84.125140, Olga}. In the zeroth order all the diagrams with unequal degree of site $\veci$ and $\vecj$ vanish, namely $I_{\veci[\vecj]\s}^{2} = I_{\veci[\vecj]}^{4} = I_{\veci\s, \vecj}^{24} = I_{\veci, \vecj\s}^{42} = T_{\veci, \vecj, \s}^{31} = T_{\veci, \vecj, \s}^{13}$. The remaining diagrammatic sums are equal to
\eqn
I_{{\veci\s} {\vecj\s}}^{22} &=& - P_{\veci \vecj\s}^2, \\
I_{{\veci\s} {\vecj\ov{\s}}}^{22} &=& S_{\veci \vecj}^2, \\ 
S_{\veci\s, \vecj\ov{\s}}^{22} &=& - P_{\veci \vecj\ua} P_{\veci \vecj\da} - S_{ij}^2, \\
T_{\veci, \vecj, \s}^{11} &=& P_{\veci \vecj\s}, \\
T_{\veci, \vecj, \s}^{33} &=& - P_{\veci \vecj\s} P_{\veci \vecj \ov{\s}}^2 - P_{\veci \vecj \ov{\s}} S_{ij}^2.
\eqnx
In this situation, and if we additionally disregard the $T^{33}$ and $I^{44}$ terms, relations valid for isotropic system are obtained
\eqn
\la \hat{c}^{\dagger}_{\veci,\sigma}\hat{c}_{\vecj,\sigma} \ra_G^{(GA)} &=& q^2 T^{11} = g_t \la \hat{c}^{\dagger}_{\veci,\sigma}\hat{c}_{\vecj,\sigma} \ra_0, \label{eq:iso1GA} \\
\la \hat{S}_\veci^{+}\hat{S}_\vecj^{-} + \hat{S}_\veci^{-}\hat{S}_\vecj^{+} \ra_G^{(GA)} &=& g_s \la \hat{S}_\veci^{+}\hat{S}_\vecj^{-} + \hat{S}_\veci^{-}\hat{S}_\vecj^{+} \ra_0,  \label{eq:iso2GA} \\
\la \hat{S}_\veci^{z}\hat{S}_\vecj^{z} \ra_G^{(GA)} &=& g_s \la \hat{S}_\veci^{z}\hat{S}_\vecj^{z} \ra_0.\label{eq:iso3GA}
\eqnx
reproducing analytically the results of GA\cite{Zhang}. \tg{It is interesting to see how big is the difference between the exact expressions, Eqs. (\ref{eq:iso1})-(\ref{eq:iso4}), and their GA approximations, Eqs. (\ref{eq:iso1GA})-(\ref{eq:iso3GA}). This difference is analyzed in Appendix \ref{app:GF}.}

If we consider general phases, and we keep the $T^{33}$ term, then the expressions for the expectation values of the hopping and the exchange term become 
\eq
\la \hat{c}_{\veci,\sigma}^{\dagger} \hat{c}_{\vecj,\sigma} \ra_G = q_{\veci, \s} q_{\vecj, \s} \left(P_{\veci \vecj\s} - P_{\veci \vecj \ov{\s}} \frac{P_{\veci \vecj\s} P_{\veci \vecj \ov{\s}} + S_{ij}^2 }{(1 - n_{\veci \ov{\s}})(1 - n_{\vecj \ov{\s}})}\right), \label{eq:hopping2}
\eqx
\eqn
\la \hat{S}_\veci^{z}\hat{S}_\vecj^{z} \ra_G &=& \frac{m_\veci m_\vecj}{4} +
   \frac{(m_\veci+1) (m_\vecj+1) (- P_{ij\da}^2)}{4 (n_{{\veci\da}}-1) (n_{{\vecj\da}}-1)}+
   \frac{(-m_\veci-1) (1-m_\vecj) S_{\veci \vecj}^2 }{4(n_{{\veci\da}}-1) (n_{{\vecj\ua}}-1)}-\nonumber \\ &&
   \frac{(-m_\veci+1)(-m_\vecj-1) S_{\veci \vecj}^2 }{4 (n_{{\veci\ua}}-1)(n_{{\vecj\da}}-1)}+
   \frac{(-m_\veci+1) (1-m_\vecj) (- P_{ij\ua}^2)}{4 (n_{{\veci\ua}}-1) (n_{{\vecj\ua}}-1)}+\nonumber \\ &&
   \frac{I_{{\veci \vecj}}^{44} m_\veci m_\vecj}{(n_{{\veci\da}}-1)(n_{{\veci\ua}}-1) (n_{{\vecj\da}}-1) (n_{{\vecj\ua}}-1)},\label{eq:SzSz2}
\eqnx
\eq
\frac{1}{2}\la \hat{S}_\veci^{+}\hat{S}_\vecj^{-} + \hat{S}_\veci^{-}\hat{S}_\vecj^{+} \ra_G = \left[(n_{{\veci\da}}-1) (n_{{\veci\ua}}-1) (n_{{\vecj\da}}-1)(n_{{\vecj\ua}}-1)\right]^{-1/2} (- P_{ij\ua} P_{ij\da} - S_{ij\s}^2). \label{eq:S+S-2}
\eqx
When the 4-line contribution from the diagrammatic sum $I_{{\veci \vecj}}^{44}$ (in Eq. (\ref{eq:SzSz2})) is neglected, our method reproduces the GCGA results \footnote{\tg{The numerical difference between the results of the zeroth order DE-GWF and GCGA is smaller than the line thickness of the presented curves, so neglecting of the diagrammatic sum $I_{{\veci \vecj}}^{44}$ is not essential.}}. Explicitly, Eqs. (\ref{eq:hopping2}), (\ref{eq:SzSz2}), and (\ref{eq:S+S-2}) are equivalent, respectively, to Eqs. (15), (20), and (21) of Ref. \onlinecite{Fukushima}. In a similar manner, the equivalence is obtained for the density-density term, Eq. (\ref{eq:density}), with the result of the GCGA approach presented in Ref. \onlinecite{Fukushima2} (Eq. (44) therein). Therefore, within the present approach the results of a sophisticated version of the RMFT \cite{Fukushima,Fukushima2} are obtained.

\subsection{Test case: one dimensional t-J model}

\begin{figure}[ht!]
\centerline{\includegraphics[angle=270,width=0.8\columnwidth]{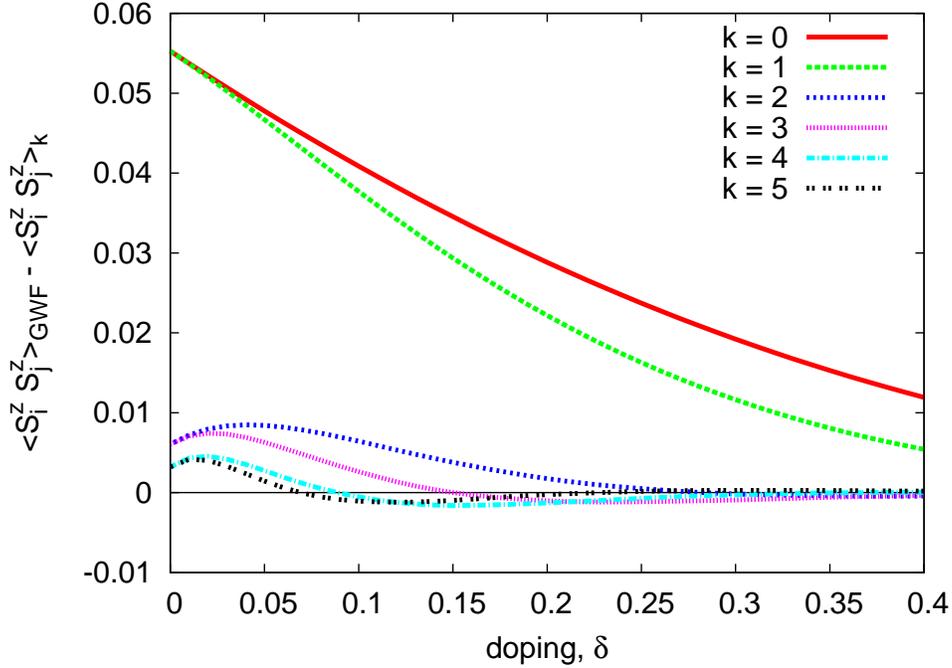}}
\caption{(Color online) Difference between the exact GWF results for the one-dimensional $t$-$J$ model and our DE-GWF results as a function of doping for orders $k=0 \div 5$. The DE-GWF results change most when an even order is taken into account (e.g., inclusion of the fourth order terms gives bigger change than inclusion of the third). The largest discrepancy of the results is close to half filling, where the expansion parameter $x$ approaches its maximal absolute value of $|x| = 4$.
\label{fig:1}}
\end{figure}

As a test case of our analytical results we consider the one-dimensional $t$-$J$ model, for which an exact analytical solution has been presented~\cite{PhysRevB.38.6911} in the paramagnetic case. We calculate the exact value of the spin-spin correlation function $\la \hat{S}_\veci^z \hat{S}_\vecj^z \ra$ using Eq. (49) from Ref. \onlinecite{PhysRevB.38.6911} and with our DE-GWF method. The difference between these two results is presented in Fig. \ref{fig:1} as a function of doping in the orders $k = 0 \div 5$. It can be seen that the fifth-order results are very close to the exact results for the doping $\delta \gtrsim 0.05$. The discrepancy should decrease with the increasing dimensionality $d$, as the zeroth order results are exact for infinite $d$. The fifth-order results are more than one order of magnitude closer to the exact values than those obtained in the zeroth order. Note also that the latter ($k=0$ results) are equivalent to those of the approach proposed in Ref. \onlinecite{Fukushima}.

The order $k$ to which we carry out our expansion, is not the only parameter affecting convergence. Another one is the number
of $\Ps$ lines \tg{(defined in Eq. (\ref{pl}))} included when calculating the diagrammatic sums. Its effect on results for the spin-spin correlation function is analyzed in Appendix \ref{app:1d}.

\section{Variational problem}

In the previous section we have provided analytical results for the expectation values of all terms appearing in the Hamiltonian (\ref{eq:H}) with respect to the assumed wave function (\ref{eq:proj2}). These results enable us to calculate the ground state energy $W \equiv \la \hat{H} \ra_G$ for a fixed $\Ps$. The remaining task is the minimization of this energy (or of the functional $\mathcal{F} \equiv W - 2 \mu_G n_G$, with $n_G \equiv \la \hat{n}_{i\s} \ra_G$) with respect to the wave function $|\Psi_0 \rangle$. This wave function enters into the variational problem via $n \equiv \la \hat{n}_{i\s} \ra_0$ and the lines $P_{\vecl,\vecl'}$ and $S_{\vecl,\vecl'}$. In the following we consider only translationally invariant wave functions. Since we study superconducting states, the correlated and non-correlated numbers of particles ($n_G$ and $n$) may differ, and hence it is technically easier to minimize the functional $\mathcal{F}$ at a constant chemical potential $\mu_G$, and not the ground state energy at a constant number of particles $n_G$. 

\begin{figure}[ht!]
\centering
\includegraphics[width=0.55\columnwidth]{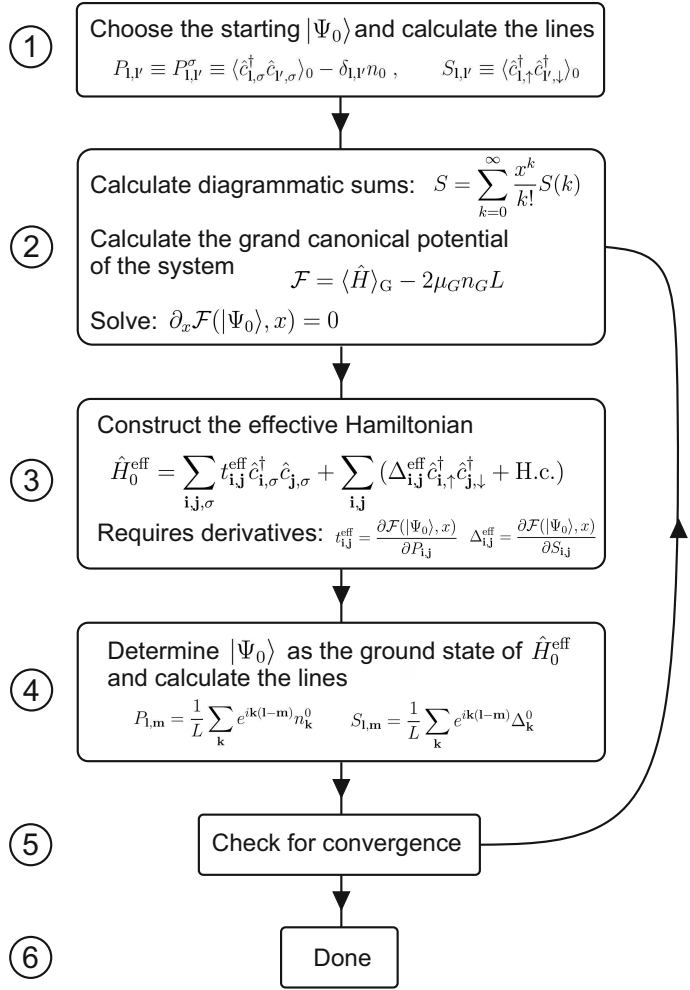}
\caption{The full self-consistency loop of the DE-GWF method.}
\label{fig:3}
\end{figure}

The remaining variational problem leads (cf. e.g. \tg{Refs. \onlinecite{Yang,Wang,PhysRevB.67.075103}}) to the effective single-particle Schr\"{o}dinger-like equation
\eq \label{eq:iou}
\hat{H}_0^{\rm eff} |\Psi_0\rangle= E^{\rm eff} |\Psi_0\rangle,
\eqx
with the self-consistently defined effective single-particle Hamiltonian
\begin{eqnarray}
\hat{H}_0^{\rm eff} &=& \sum_{\veci,\vecj, \s}t^{\rm eff}_{\veci,\vecj}\label{eq:iou0}
\hat{c}_{\veci,\sigma}^{\dagger}\hat{c}_{\vecj,\sigma}^{\phantom{\dagger}} + \sum_{\veci,\vecj} \big( \Delta^{\rm eff}_{\veci,\vecj} \hat{c}_{\veci,\uparrow}^{\dagger}\hat{c}_{\vecj,\downarrow}^{\dagger} + {\rm H.c.} \big),\\ \label{eq:iou1}
t^{\rm eff}_{\veci,\vecj} &=& \frac{\partial \mathcal{F} (|\Psi_0\rangle,x)}{\partial P_{\veci,\vecj}} \;, \qquad
\Delta^{\rm eff}_{\veci,\vecj} = \frac{\partial \mathcal{F} (|\Psi_0\rangle,x)}{\partial S_{\veci,\vecj}} \;.  \label{eq:iouF}
\end{eqnarray}
\tg{The effective dispersion relation, the effective gap, and eigenenergies of $\hat{H}_0^{\rm eff}$ are defined as}
\begin{eqnarray}
\varepsilon^{\rm eff}(\veck) &=& \frac{1}{L}\sum_{\veci,\vecj}\exp^{{\rm i}(\veci-\vecj)\veck} t^{\rm eff}_{\veci,\vecj} = \left[ \sum_{\vecj}\exp^{{\rm i}(\veci-\vecj)\veck} t^{\rm eff}_{\veci,\vecj} \right]_{\veci = (0,0)}, \\
\label{eq:effedispersion}
\Delta^{\rm eff}(\veck) &=& \frac{1}{L}\sum_{\veci,\vecj}\exp^{{\rm i}(\veci-\vecj)\veck} \Delta^{\rm eff}_{\veci,\vecj} = \left[ \sum_{\vecj}\exp^{{\rm i}(\veci-\vecj)\veck} \Delta^{\rm eff}_{\veci,\vecj} \right]_{\veci = (0,0)}, \\
E^{\rm eff}(\veck) &=& \sqrt{\varepsilon^{\rm eff}(\veck)^2 + \Delta^{\rm eff}(\veck)^2},
\end{eqnarray}
respectively, where the last expressions for $\varepsilon^{\rm eff}(\veck)$ and $\Delta^{\rm eff}(\veck)$ are valid for a homogeneous system.
The final solution (of one iteration of our self-consistency loop) is obtained by solving Eqs.~(\ref{eq:iou})-(\ref{eq:iouF}), with the additional minimization condition, $\partial_x \mathcal{F}(|\Psi_0\rangle, x) = 0$. Having solved these equations, we can make the next iteration and calculate the new $|\Psi_0\rangle$ lines \tb{(from definition in Eq. (\ref{pl}))}, according to the prescriptions
\eqn
P_{\vecl, \vecm} &=& \frac{1}{L} \sum_\bk e^{i \bk (\vecl - \vecm)} n_{\bk}^0, \qquad
n_{\bk}^0 = \frac{1}{2} \left[ 1 - \frac{\varepsilon^{\rm eff}(\veck)}{E^{\rm eff}(\veck)} \right],
\label{eq:PLine} \\
S_{\vecl, \vecm} &=& \frac{1}{L} \sum_\bk e^{i \bk (\vecl - \vecm)} \Delta_{\bk}^0, \qquad
\Delta_{\bk}^0 = \frac{1}{2} \frac{\Delta^{\rm eff}(\veck)}{E^{\rm eff}(\veck) }.
\label{eq:SLine} \\
\eqnx
%
The resulting self-consistency loop is shown in Fig. \ref{fig:3}. The convergence is achieved when the new $|\Psi_0\rangle$ lines differ from the previous ones by less than the assumed precision value, typically $10^{-7}$.

\section{Results}

\begin{figure}[ht!]
\centerline{\includegraphics[width=0.5\columnwidth,angle=270]{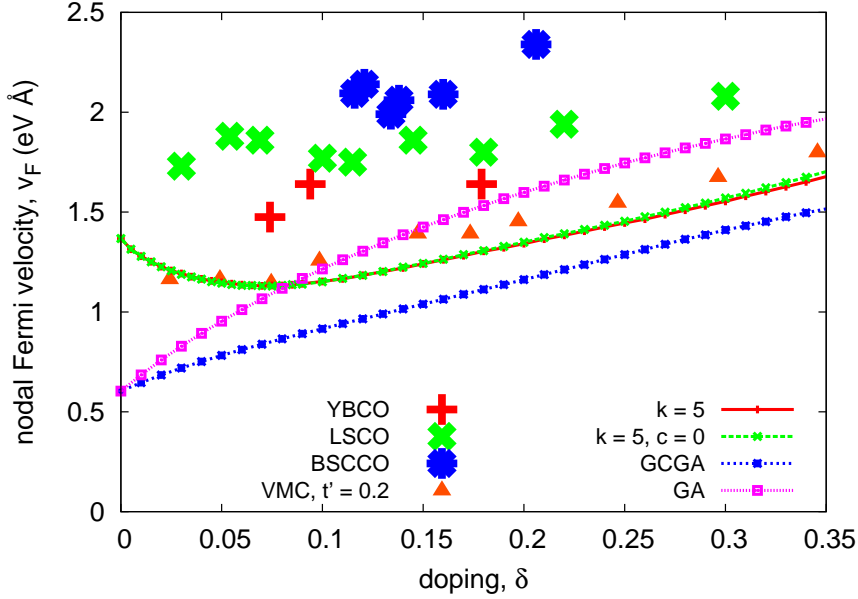}}
\caption{(color online) Universal Fermi velocity in the nodal ($k_x = k_y$) direction as a function of doping. The experimental values are taken from Ref.~\onlinecite{EdeggerPRL} and references therein and have typically an uncertainty of $20\%$. The VMC results are taken from Ref.~\onlinecite{Yunoki}. The results obtained for the model with and without the density-density term (for $c=0$ and $c=1$, respectively) are very close.
\label{fig:5}}
\end{figure}

\begin{figure}[ht!]
\includegraphics[width=0.55\columnwidth]{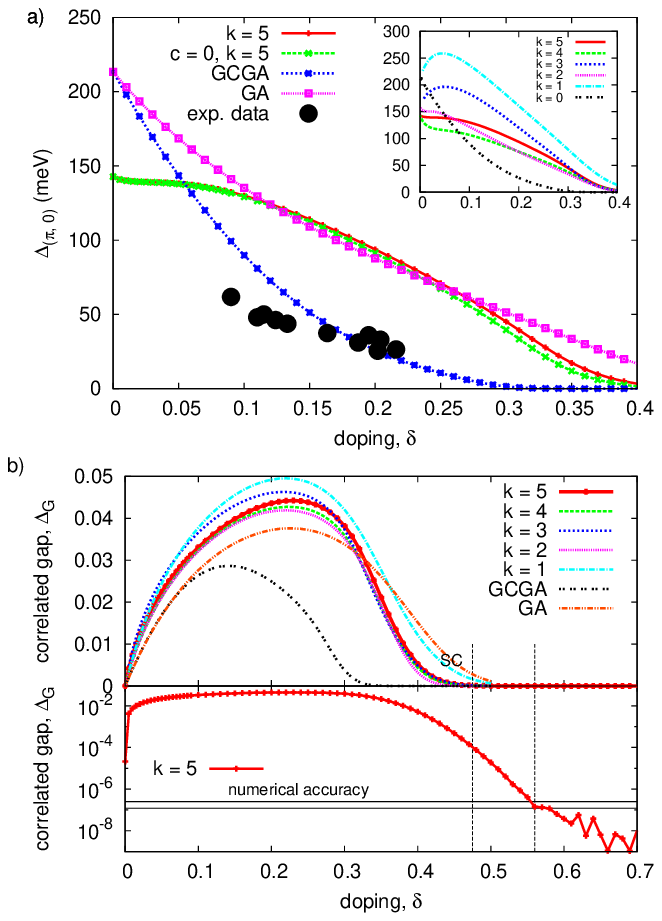}
\caption{(Color online) (a) the effective gap at the antinodal point $\Delta^{\rm eff}_{\bk = (\pi, 0)}$ and its comparison with the experimental data of Ref. \onlinecite{PhysRevLett.83.3709}; (b) the correlated gap $\Delta_G$. In (a) the gap values are plotted in physical units (assuming $t=0.35\, {\rm eV}$). The zeroth to fifth order results are exhibited to demonstrate the method convergence. In (b) we show also the gap on the logarithmic scale in the lower panel. We mark our numerical accuracy limit, which is around $10^{-7}$, by two horizontal lines. A residual very small gap persists to the dopings beyond the upper critical concentration (see also main text).}
\label{fig:6}
\end{figure}

The self-consistency loop in Fig. \ref{fig:3} is solved numerically with the use of GNU Scientific Library (GSL). The new lines are calculated from Eqs. (\ref{eq:PLine})-(\ref{eq:SLine}) by numerical integration in $\bk$ space (\tg{this corresponds to an infinite system size, $L \to \infty$}). The typical accuracy of our solution procedure is equal to $10^{-7}$. We set $|t| = -t$ as our unit of energy and, unless stated otherwise, and present the results for $t'=0.25$ and $J=0.3$. We consider the two cases with $c=0$ and $c=1$, but, as their results are very close, we show the $c=0$ data only in Figs. \ref{fig:5} and \ref{fig:6}a. In several figures we provide also the results of the GCGA (and GA) methods, which were obtained by the simplified zeroth order DE-GWF method (equivalent to GCGA or GA, as discussed in Sec. \ref{sec:equiv}).

We carry out the expansion to the fifth order, which in most cases provides quite accurate results. The lower-order results are also exhibited in selected figures to visualize our method's convergence. To calculate the diagrammatic sums we need to neglect long-range $\Ps$-lines in real space. Namely, we take as nonzero only the lines $P_{\veci, \vecj} \equiv P_{0, (\veci-\vecj)} \equiv P_{XY}$ (with $X = i_1 - j_1$, $Y = i_2-j_2$), for which $X^2 + Y^2 \leq R^2 = 25$ (i.e., with 14 neighbors). The same condition applies for $S_{\veci, \vecj}$, $t^{\rm eff}_{\veci, \vecj}$, and $\Delta^{\rm eff}_{\veci, \vecj}$. We also define an additional convergence parameter. Namely, we take into account only those contributions to the diagrammatic sums, in which the total Manhattan distance (i.e., $|X|+|Y|$) of all lines is smaller than $R_{\rm tot}$ typically set to $R_{\rm tot} = 26$.

In total, we have the three convergence parameters: (i) order $k$, (ii) $\Ps$ cutoff radius $R$, (iii) total Manhattan distance of all lines $R_{\rm tot}$. The uncertainty of our results coming from parameters (ii) and (iii) is of the order of line thickness of the presented curves, whereas the $k$-th order results for most doping values are between the $k-1$ and $k-2$ order results (and the differences between them diminish with the increasing $k$). Therefore, we believe that the series is convergent. \tg{The accuracy of our results may be further improved by including higher order terms. However, in the sixth order there are already $10^7$ nonequivalent SC diagrams for the $T^{33}$ diagrammatic sum, what makes the analysis computationally demanding. Alternatively, as in the bold diagrammatic Monte Carlo technique \cite{PhysRevB.77.125101,Houcke}, a Ces\`{a}ro-Riesz summation method could be used (cf. Ref. \onlinecite{PhysRevB.77.125101}, Sec. V.) to improve convergence of the diagrammatic sums. Work along these lines is planned.}

\begin{figure}[ht!]
\includegraphics[width=0.55\columnwidth]{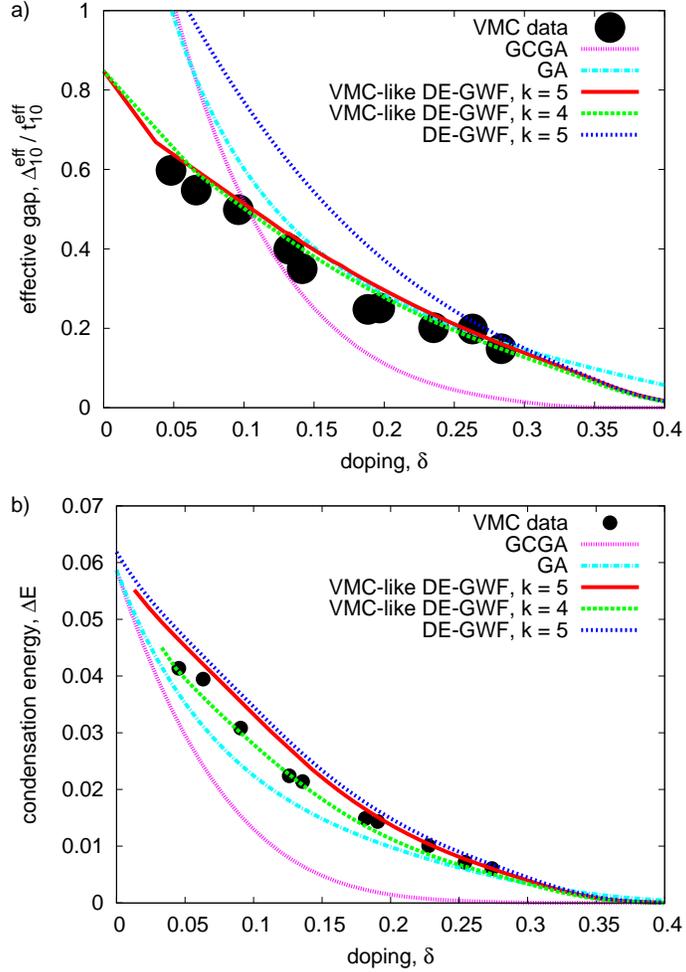}
\caption{(Color online) Comparison of DE-GWF (lines) for $J=0.3$ and $t' = 0.3$ with variational Monte Carlo (VMC) results (the point size is equal to the error; from Ref. \onlinecite{PhysRevB.74.165109}). (a) Effective gap (in units of $t_{10}^{\rm eff}$) and (b) condensation energy as a function of doping. The VMC-like DE-GWF lines are obtained with effective single-particle Hamiltonian containing only (next)nearest-neighbor and on-site terms (see Appendix \ref{app:B} for details). The fourth and fifth order results are shown for VMC-like DE-GWF \tg{to illustrate the convergence}. The GA and GCGA results are obtained by the zeroth order DE-GWF.}
\label{fig:4}
\end{figure}

To test our approach, in Figs. \ref{fig:4}ab we have compared our results with those of Ref. \onlinecite{PhysRevB.74.165109} obtained by variational Monte Carlo (VMC) method for the Hamiltonian with $c=1$ and for the values of parameters $t' = J = 0.3$. In order to obtain comparable results we have to truncate our effective Hamiltonian, as in VMC, so that it contains parameters only up to next nearest-neighbors (see Appendix \ref{app:B} for details). We call the resulting approach VMC-like DE-GWF. Its results agree very well with those of VMC. The sources of small quantitative discrepancies between the two results are due to approximations of both methods. First, in VMC calculations, a finite-size $11\times11$ (or $13\times13$) lattice is used, whereas we use an infinite lattice in the DE-GWF method. Note that in an analogous comparison\cite{PhysRevB.88.115127} with VMC calculations performed for the Hubbard model on an $8 \times 8$ lattice, the discrepancies were much larger. Second, in our method we perform the expansion up to the 5th order (the remaining error coming from the $\Ps$ cutoff in real space is of the order of line thickness).

Additionally, discrepancies might come from the fact that in our procedure the correlated ($n_G$) and uncorrelated ($n$) numbers of particles are slightly different, whereas it is not clear to us from Ref.~\onlinecite{PhysRevB.74.165109} if there is a change in the particle number there due to the Gutzwiller projection.

The difference between the VMC-like DE-GWF and the full DE-GWF scheme shows that neglecting longer-range gap and hopping components can lead to a \tg{decrease of the principal gap component by up to $75\%$ (the largest discrepancy is near the half filling) and corresponding decrease of the condensation energy by $3 \div 35\%$ (the largest discrepancy is for overdoped system)}. These discrepancies are larger than those observed in Ref. \onlinecite{Watanabe}, in which the longer-range \textit{hopping} components were not included. Our results suggest that inclusion of the longer-range effective parameters is important as it can lead to changes of results even by a factor of 1.75, even though the condensation energy does not change much. We also provide GA and GCGA results to show their qualitative differences with respect to both VMC and DE-GWF. \tg{Surprisingly, GA is closer to the VMC and the DE-GWF results than its improved variant, GCGA. The largest discrepancy between GA and either VMC or DE-GWF data is for underdoped ($\delta < 0.15$) and overdoped $\delta>0.3$ systems.} We have also verified that for the zeroth order the VMC-like and the full DE-GWF methods yield the same results, as it should be, because the zeroth order diagrammatic sums only contain lines connecting (next)nearest neighbors.

\tg{The break in the VMC-like DE-GWF curve in Fig.~\ref{fig:4}a appearing at $\delta \approx 4 \%$ is related to the phase separation effect present for the SC phase in the $t$-$J$ model in both (VMC-like)DE-GWF and VMC methods \cite{PhysRevB.70.104503, PhysRevB.76.140505, PhysRevB.85.104511}. Namely, the chemical potential ($\mu_G$) of the SC phase has a maximum as a function of doping for $\delta \approx 3 \div 5 \%$. For this reason, our numerical procedure (in which $\mu_G$ is increased at each step) fails to converge for $\delta \lesssim 5 \%$. To obtain the following DE-GWF results we changed our method to work with a fixed $n_G$ (similarly, as in Refs. \onlinecite{PhysRevB.84.125140,Olga,Zegrodnik,Abram}, with an additional equation for $n_G$). This allowed us to obtain convergence in the vicinity of half filling.}

One of the most important physical characteristics of the cuprates is the universal nodal Fermi velocity $v_F$ (i.e., $v_F$ is independent of $\delta$)~\cite{Zhou}. Recently, it has been shown however that the Fermi velocity for the underdoped samples exhibits a low-energy kink and a nontrivial doping dependence~\cite{Vishik2}. The velocity posesses the two components: one near the Fermi surface which is doping dependent and the velocity slightly below the Fermi surface which is doping independent. The source of the kink in the dispersion is probably the electron-phonon interaction~\cite{Johnston} and is not included in our purely electronic model.
In Fig. \ref{fig:5} we show the Fermi velocity defined as $v_F = \nabla_\bk \epsilon^{\rm eff}(\bk)|_{\epsilon^{\rm eff}(\bk) = 0}$. Its behavior agrees with the experimental results (we assume the lattice constant $a=4\, {\rm \AA}$ and $|t|=0.35\, {\rm eV}$). The RMFT method does not reproduce such behavior~\cite{Jedrak2,EdeggerPRL}. We also present for comparison the \tb{VMC results~\cite{Yunoki,Paramekanti,PhysRevB.69.144509,PhysRevB.70.054504} obtained in Ref. \onlinecite{Yunoki} for $t' = 0.2$}. The weak doping dependence of the Fermi velocity speaks in favor of a transfer of the spectral density to the nodal direction from the antinodal direction with the decreasing doping (see also the discussion below).

In Fig. \ref{fig:6} we plot the two gaps: the effective gap at the antinodal point $\Delta^{\rm eff}_{\bk = (\pi, 0)}$ and the correlated gap $\Delta_G$. The effective gap agrees with the experimental values only after rescaling by a factor of $0.4$ (not shown) similarly as for the \tb{GA~\cite{EdeggerPRL} and VMC~\cite{PhysRevB.70.054504} approaches}. \tg{Recent experiments have shown however, that the competition between the superconducting gap and pseudogap \cite{Sacuto,Shekhter,Efetov} in BSCCO diminishes essentially the value of the superconducting gap in the nodal direction \cite{KondoT,KondoT2,PhysRevLett.101.227002}. In fact, this gap is shown to vanish for underdoped samples \cite{KondoT}. Therefore, a quantitative agreement with the experimental points in Fig. \ref{fig:6}a should not be the goal in describing high-temperature superconductors, as including the pseudogap may change the picture essentially.} One should also keep in mind that $\Delta^{\rm eff}_{\bk = (\pi, 0)}$ depends on $J$. For lower $J$ values we obtain much better agreement with the experiment (but at the same time, the agreement of the nodal Fermi velocity is then worse). Similarly as in VMC calculations~\cite{Yokoyama2}, we observe an exponential decay of the gap with the doping reaching the upper critical concentration \tg{$\delta_c \sim 1/2$}. \tg{We term as SC the phase with $\Delta_G > 10^{-4}$, which corresponds to gap values of the order of $0.4 \, {\rm K}$, below which other effects can destabilize the superconductivity.} In our model situation however, we still have a stable superconducting solution even if we increase doping above such defined $\delta_c$ by $8\%$. One must note that if the experimentally measured gap is usually determined for temperature $T \gtrsim 1 {\rm K}$, then the tail of $\Delta_G(\delta)$ beyond $\delta_c$ will not be detected as then effectively $T>T_c$. In the inset of Fig. \ref{fig:6}a and in the upper panel of Fig. \ref{fig:6}b we show also the order-of-expansion dependence of the results. It can be seen that, for most of the doping values, the $k$-th order results are between the results obtained for order $k-1$ and $k-2$. Moreover, the difference between the orders diminishes with the increasing order.

\begin{figure}[ht!]
\includegraphics[width=0.75\columnwidth]{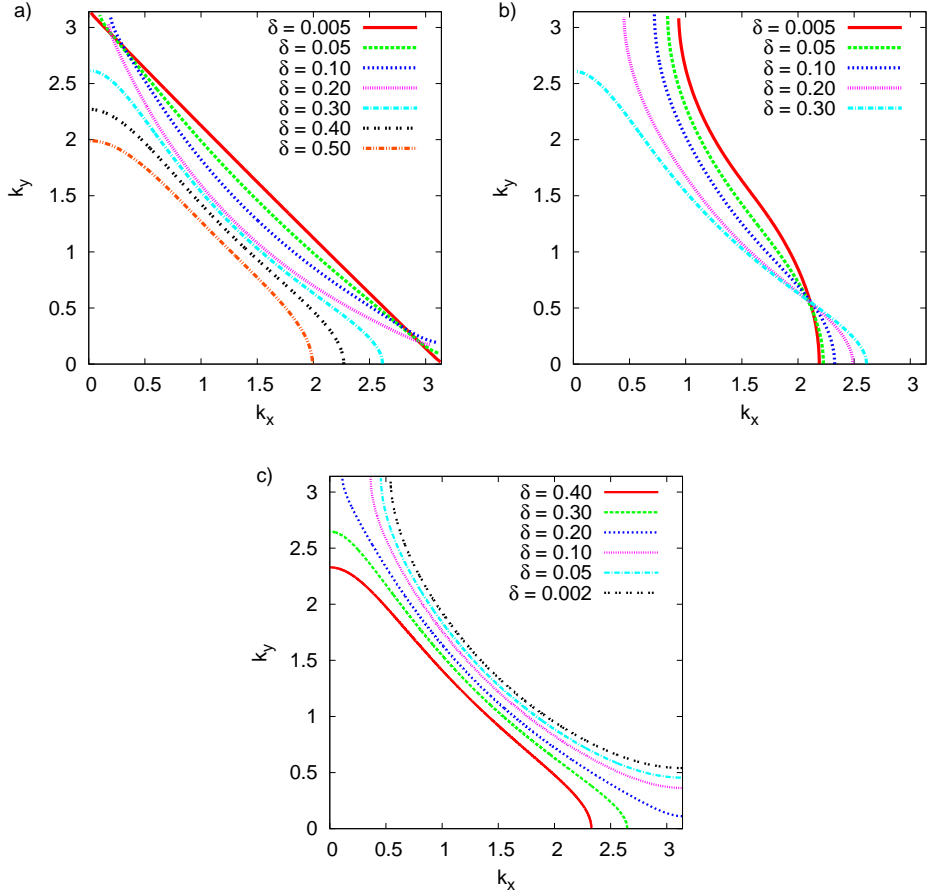}
\caption{(Color online) (a) Fermi surfaces for selected values of doping in the normal phase (a), the spontaneously-distorted Pomeranchuk phase (b), and for the bare Hamiltonian with only kinetic energy without renormalization (c).}
\label{fig:7a}
\end{figure}

\begin{figure}[ht!]
\includegraphics[width=0.5\columnwidth]{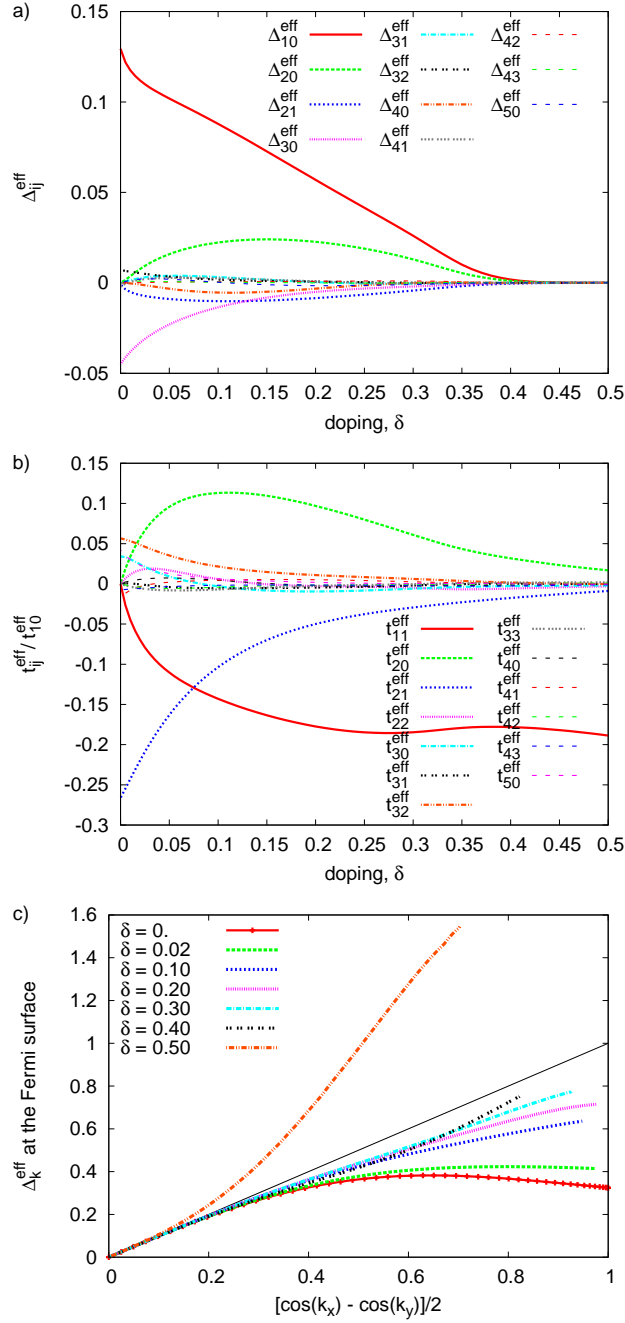}
\caption{(Color online) (a) Effective gap parameters obtained variationally as a function of doping; (b) effective hopping parameters relative to the dominant $t_{10}^{\rm eff}$ contribution; (c) effective gap in momentum space at the Fermi energy for selected doping values. The black line corresponds to a pure $d_{x^2 - y^2}$ dependence. The gaps are normalized \tg{for clarity}.}
\label{fig:8}
\end{figure}

In the panel composing Fig. \ref{fig:7a} we exhibit the doping dependence of the Fermi-surface topology, starting from the effective Hamiltonian (\ref{eq:iou0}). We also show results for the state with a spontaneously broken rotational symmetry, i.e., the appearance of the so-called Pomeranchuk phase \cite{JPSJ.69.2151,*JPSJ.69.332,PhysRevLett.85.5162,Jedrak1}. \tg{This phase has also been investigated by VMC \cite{PhysRevB.74.165109,Zheng}. The drawback of using VMC in such calculations is that the finite-size effects become much more important than for the description of the SC phase (typically $12 \times 12$ points~\cite{Zheng} or $8 \times 8$ points \cite{PhysRevB.74.165109} are included within the quarter of the Brillouin zone, cf. also the discussion in Ref.~\onlinecite{PhysRevB.74.165109}). Our method does not suffer from those finite-size limitations and therefore, it seems \tg{more} appropriate for analyzing the Fermi-surface properties.} It can be seen from Fig. \ref{fig:7a}b that the correlated Fermi surface differs essentially from the non-interacting one near half filling. Namely, if we approach the half-filled case the Fermi surface becomes a line as in a bare Hamiltonian with the n.n. hopping only. This is caused by diminishing of certain effective hopping parameters in the vicinity of the half filling (as shown explicitly in Fig. \ref{fig:8}b below). The doping dependence of the Fermi surface in the Pomeranchuk phase is similar to that obtained in the Hubbard model \cite{Buenemann}. The role of the Pomeranchuk instability will not be studied in detail here.

\begin{figure}[ht!]
\includegraphics[width=0.5\columnwidth]{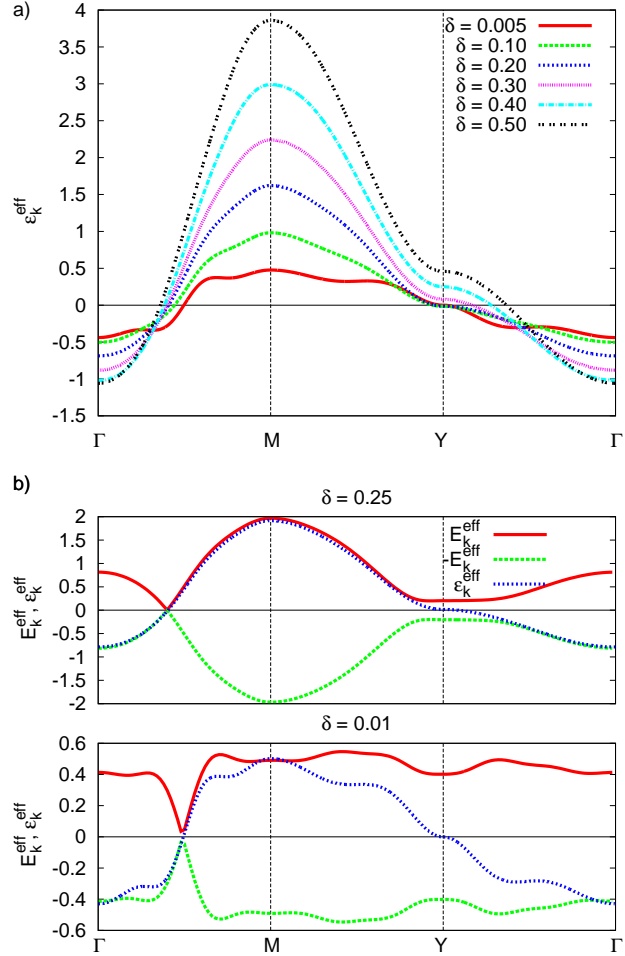}
\caption{(Color online) (a) Dispersion relation of the effective Hamiltonian for the normal (paramagnetic) phase. The vertical lines mark specific points of the Brillouin zone: $\Gamma = (0, 0 )$, $M = (\pi, \pi)$, and $Y = (0, \pi)$. The horizontal line at $\varepsilon^{\rm eff}_\bk = 0$ marks the Fermi energy. (b) SC-phase quasiparticle energies for two doping values; the energies $\varepsilon_\bk^{\rm eff}$ (of the normal phase at the same doping) are drawn for comparison.}
\label{fig:7}
\end{figure}

The dispersion relation in the normal phase and the quasiparticle energies in the superconducting state are shown in Fig. \ref{fig:7}. With the decreasing doping the bandwidth becomes smaller, and the dispersion deviates significantly from the simple form with the dominating n.n. hopping.
The SC-phase quasiparticle energies (shown in Fig. \ref{fig:7}b) resemble the metallic dispersion $\varepsilon^{\rm eff}_\bk$ only for substantial doping values. With the decreasing doping deviations coming from the effective gap become larger. This effective gap has its maximum value (in the antinodal direction) close to the $Y$ point of the Brillouin zone. For small doping values this gap is comparable to the maximum value of $\varepsilon^{\rm eff}_\bk$.

In the panel composing Fig. \ref{fig:8} we detail the effective gap and the effective hopping amplitudes. Near half filling, only a few components of the gap are of substantial magnitude, namely $\Delta^{\rm eff}_{10}$, $\Delta^{\rm eff}_{21}$ (small, but nonzero), $\Delta^{\rm eff}_{30}$, and $\Delta^{\rm eff}_{32}$, as also found in Ref. \onlinecite{Watanabe} (in which $[31]$ and $[40]$ are the most distant gap components). The same components of the effective \textit{hopping} are nonzero at half filling, together with additional ones (e.g. $t^{\rm eff}_{50}$). From Fig.~\ref{fig:8}c it follows that the effective gap along the Fermi surface deviates from a pure $d_{x^2 - y^2}$ form, \tg{especially close to half-filling, for which the gap in the antinodal direction is diminished by a factor of $3$ with respect to the pure $d_{x^2 - y^2}$ form. Such deviations are also observed in high-temperature superconductors~\cite{Mesot,Yoshida,Lee,Vishik,Vishik3,KondoT,KondoT2,PhysRevLett.101.227002}, where the situation is complicated further by the appearance of pseudogap \cite{Shekhter,Efetov,PhysRevB.85.054509}. Namely, for the underdoped samples the \textit{total} gap measured in angle-resolved photoemission spectroscopy (ARPES) is increased in the antinodal direction with respect to the pure $d_{x^2 - y^2}$ component~\cite{Mesot,Yoshida,Lee,Vishik,Vishik3}, but the spectral weight corresponding to the \textit{superconducting gap} is simultaneously decreased \cite{KondoT,KondoT2,PhysRevLett.101.227002}, which agrees with our findings in Fig.~\ref{fig:8}c. Therefore, such decrease of superconducting gap can be an intrinsic effect for strongly-correlated superconductors, not only caused by the competition with the pseudogap.}

\section{Summary and Outlook}

\subsection{Methods comparison}

\tg{When working with a variational Gutzwiller wave function, the main task is the calculation of the expectation value (Eq. (\ref{eq:W})) of the Hamiltonian with respect to this trial wave function. So far, there have been two types of methods to approach this problem. In one of them (GA and the derivatives) the expectation values of the Hamiltonian terms are approximated by the corresponding expectation values with respect to the non-correlated wave function ($\Ps$) multiplied by appropriate renormalization factors (e.g. $\la \hat{c}^{\dagger}_{\veci,\sigma}\hat{c}_{\vecj,\sigma} \ra_G^{(GA)} = g_t \la \hat{c}^{\dagger}_{\veci,\sigma}\hat{c}_{\vecj,\sigma} \ra_0$). This yields a very fast method, but constitutes an additional approximation, which prevents the description of superconductivity or Pomeranchuk phase in the Hubbard model. In contrast, VMC evaluates the expectation values in a controlled manner, but on a finite lattice, which leads to an increased numerical complexity of the approach. In DE-GWF the averages are also calculated as accurately as possible, but a different path towards computing them is undertaken. The resultant procedure leads to principal advantages over VMC: (i) the absence of the finite-size limitations, (ii) relatively low computational complexity, (iii) the ability to account for longer-range effective parameters in a natural manner (iv) the possibility of extending the approach to nonzero temperatures. On the other hand, VMC can be easily extended by introducing additional Jastrow factors to the trial wave function (this yields wavefunctions with, e.g., the doublon-holon correlation \cite{Yokoyama1,Yokoyama2} or Baeriswyl wavefunctions \cite{Baeriswyl,Hetenyi,Eichenberger}). Investigation of the possibility of extending the DE-GWF method in this direction is planned.}

\subsection{Comparison with the Hubbard model results and the experiment}

As the paper contains a new method of approach (DE-GWF) to high-temperature superconductivity analyzed within the $t$-$J$ model, a methodological note is in place here. Namely, we would like to relate the present results to those coming from our very recent analysis of the Hubbard model within DE-GWF \cite{PhysRevB.88.115127}. First, the ``dome-like'' shape of $\Delta_G(\delta)$ is similar in both situations, particularly in the large-$U$ limit for the Hubbard model, though the upper critical concentration is lower in the latter case. Second, the doping independence of the Fermi velocity $v_F(\delta)$, representing a crucial test for any theory, is also closer to the experimental values in the Hubbard-model case. In both situations, DE-GWF provides much better values than those obtained within the dynamic mean-field theory (DMFT) \cite{Civelli1}. Third, the doping dependence of the gap in the antinodal direction (cf. Fig. \ref{fig:6}a) can reproduce the experimental trend if we rescale the results by a factor of $\sim 1/2$ (see also below). \tg{Fourth, the deviations of the gap value along the Fermi surface from the $d_{x^2-y^2}$-wave symmetry are consistent with the experimental trend: diminishing of the superconducting gap in the antinodal region for underdoped samples.}

\subsection{Outlook}

Combining the above features, together with a good agreement of the present results with the VMC analysis for small systems, DE-GWF provides a unique method of accounting for the basic superconducting properties in a quantitative manner. However, it fails to address one principal property, namely the appearance of the pseudogap. Very recently, we have generalized the analysis of the projected $t$-$J$ model \cite{JSunpublished} by introducing in a systematic manner its supersymmetric (spin-fermion) representation. In this new representation, the Fermi sector provides essentially the $t$-$J$ model in the above fermionic representation, with an additional renormalization of both the hopping and the kinetic exchange integral amplitudes. This should diminish the scale of energies obtained theoretically in Fig. \ref{fig:6}a (this idea requires still a detailed numerical analysis). What is even more interesting, the newest model\cite{JSunpublished} provides an explicit pairing and a separate scale of excitations in the Bose sector which may be interpreted as an appearance of a pseudogap. Summarizing, the new model preserves essential features of the $t$-$J$ model as discussed here, but introduces additionally the bosonic branch of collective phenomena. Such division is implicit in the recent calculations \cite{PhysRevB.85.054509}. Work along this line is in progress and, as it requires a very complex numerical analysis, will be presented separately in the near future.

In conclusion, it is in our view rewarding that the Hubbard and the $t$-$J$ models provide converging results, at least on a semiquantitative level. To what extent this analysis can be enriched on the same level by a multiband model \cite{Hanke1}, remains to be seen.

\begin{acknowledgments}
The work was supported in part by the Foundation for Polish Science (FNP) under the `TEAM' program, as well as by the project `MAESTRO' from National Science Centre (NCN), No.\ DEC-2012/04/A/ST3/003420. One of the authors (JK) acknowledges the hospitality of the Leibniz Universit\"{a}t in Hannover during the finalization of the paper, whereas JB thanks Jagiellonian University for its hospitality during the early stage of the work.
\end{acknowledgments}

\appendix

\section{Exemplary types of diagrams} \label{app:A0}

\begin{figure}[ht!]
\centering
\includegraphics[width=0.95\columnwidth]{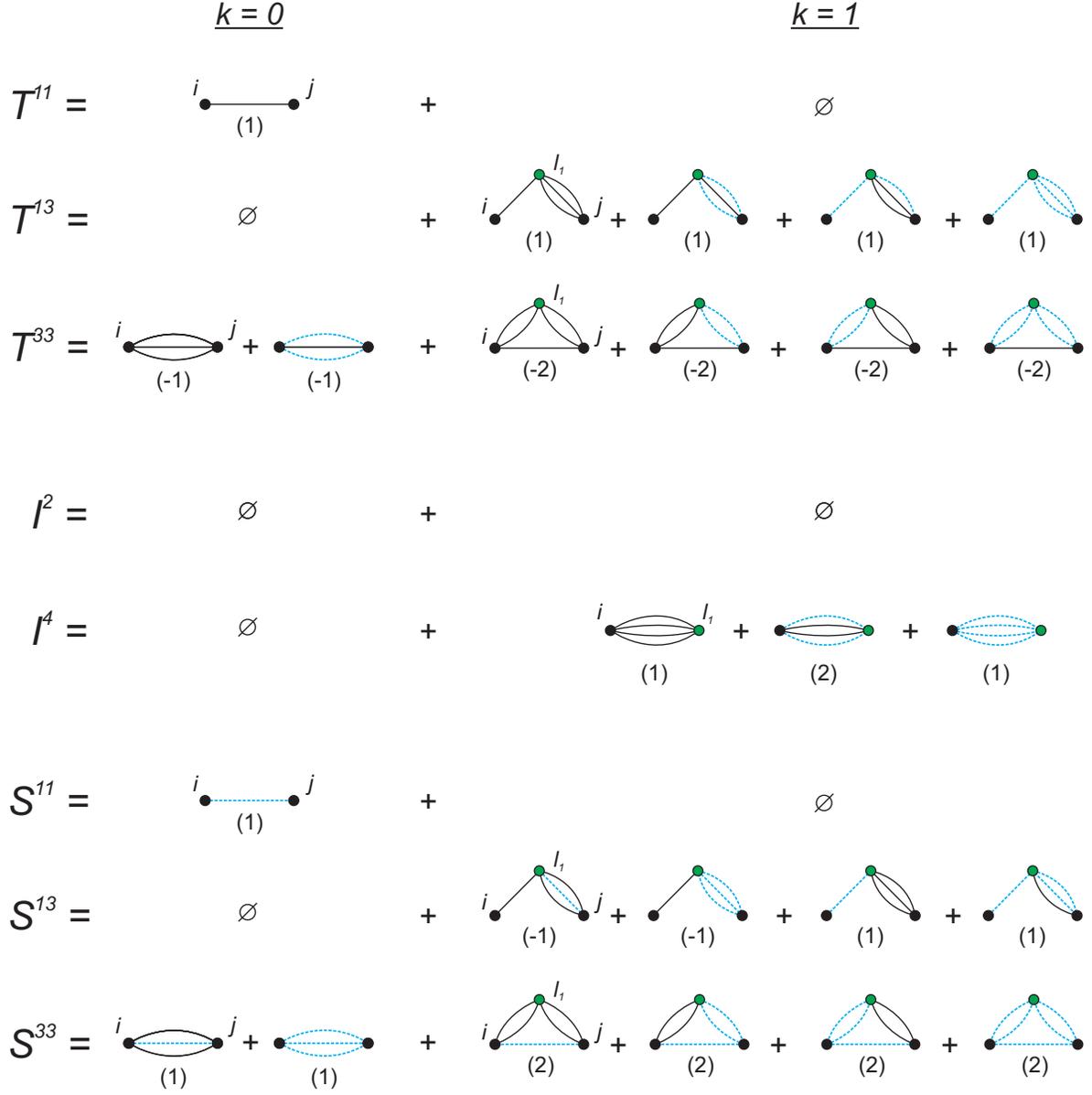}
\caption{(color online). Diagrams in the zeroth and the first order. The superconducting (paramagnetic) contractions $S_{l, l'}$ ($P_{l, l'}$) are marked with dashed (solid) lines. The internal (external) vertices are marked with green (black) circles. The numbers in brackets below diagrams represent their multiplicity (a combinatorial factor).}
\label{fig:S1}
\end{figure}

In Fig. \ref{fig:S1} we present the diagram types for the \tg{kinetic energy ($T^{11}$, $T^{13}$, $T^{33}$), the potential energy ($I^{2}$, $I^{4}$), and the ``correlated delta'' ($S^{11}$, $S^{13}$, $S^{33}$) diagrammatic sums. We consider the first two orders (i.e., the diagrams with zero and one internal vertex).} For the paramagnetic phase we would have only the diagrams without dashed lines (and obviously, no correlated delta diagrams). \tb{The number of diagrams grows exponentially with the increasing order $k$, and therefore we determine these diagrams by means of a numerical procedure.}

The general form of the resulting diagrammatic sums is obtained as e.g.
\eqn
T^{11} &=& P_{ij} + \mathcal{O}(x^2),  \label{eq:T11g} \\
T^{13} &=& x \sum_{l_1} \Big( P_{il_1}P_{jl_1}^3 + P_{il_1}P_{jl_1}S_{jl_1}^2 + S_{il_1}P_{jl_1}^2S_{jl_1} + S_{il_1}S_{jl_1}^3 \Big)+ \mathcal{O}(x^2), \\
S^{13} &=& x \sum_{l_1} \Big( - P_{il_1}P_{jl_1}^2S_{jl_1} - P_{il_1}S_{jl_1}^3 + S_{il_1}P_{jl_1}^3 + S_{il_1}S_{jl_1}^2P_{jl_1} \Big)+ \mathcal{O}(x^2). \label{eq:T13g}
\eqnx
In order to perform the summations of diagrams over a lattice, we need to assume as nonzero the $|\Psi_0\rangle$ lines up to some finite distance. In the main text we have taken as nonzero the lines ($S_{\veci, \vecj} \equiv S_{X,Y}$ with $X = (i_1 - j_1), Y = (i_2 - j_2)$, $P_{X, Y}$ - analogously) fulfilling \tg{$X^2 + Y^2 \leq R^2 = 25$. If the cutoff distance is defined by $X^2 + Y^2 \leq 2$, then they are as follows}
\eqn
T^{11}(0) &=& P_{10}, \\
T^{33}(0) &=& - P_{10}^3 - P_{10} S_{10}^2, \\
T^{13}(1) &=& 2 P_{10}^3 P_{11} + 2 P_{10} P_{11}^3 + 2 P_{10} P_{11} S_{10}^2.
\eqnx
Increasing the \tg{cutoff distance $R$} leads to significant complication of the obtained expressions - e.g. for $X^2 + Y^2 \leq 4$ (allowing for nonzero $P_{20}$ and $S_{20}$ lines), we have
\eqn
T^{13}(1) &=& 2 P_{10}^3 P_{11} + 2 P_{10} P_{11}^3 + 2 P_{10} P_{11} S_{10}^2 + P_{10}^3 P_{20} + S_{10} S_{20}^3 + P_{10}^2 S_{10} S_{20} \nonumber \\
&& + P_{20}^2 S_{10} S_{20} + 2 P_{10} P_{11} S_{10}^2 + S_{10}^3 S_{20} + P_{10} P_{20} S_{20}^2.
\eqnx
In our numerical procedure, when calculating the diagrammatic sums, we start from the general form (as in Eqs. (\ref{eq:T11g}) - (\ref{eq:T13g})) and sum over the internal vertices positions (here over $\vecl_1$) making sure that the condition $X^2 + Y^2 \leq 25$ is fulfilled for all contributing lines $P_{X,Y}$, $S_{X,Y}$.

\section{Exchange term evaluation} \label{app:A}

\tg{The expressions for the components of the exchange term are as follows (with $m_i \equiv n_{i\ua} - n_{i\da}$)}

\eqn
\label{eq:density}
\frac{1}{4} \la \hat{n}_\veci \hat{n}_\vecj \ra_G &=& +\frac{n_\veci n_\vecj}{4} +
 \frac{(n_\veci-1) (n_\vecj-1) I_{\veci {\vecj\da}}^{42}}{2 (n_{{\veci\da}}-1) (n_{{\veci\ua}}-1) (n_{{\vecj\da}}-1)}+
\frac{(n_\veci-1) (n_\vecj-1) I_{\veci{\vecj\ua}}^{42}}{2 (n_{{\veci\da}}-1) (n_{{\veci\ua}}-1)(n_{{\vecj\ua}}-1)}+\nonumber \\ &&
\frac{(n_\veci-1) (n_\vecj-1) I_{{\veci\da} \vecj}^{24}}{2 (n_{{\veci\da}}-1) (n_{{\vecj\da}}-1) (n_{{\vecj\ua}}-1)}+
\frac{(n_\veci-1) (n_\vecj-1) I_{{\veci\da} {\vecj\da}}^{22}}{4 (n_{{\veci\da}}-1)(n_{{\vecj\da}}-1)}+
\frac{(n_\veci-1) (n_\vecj-1) I_{{\veci\da} {\vecj\ua}}^{22}}{4 (n_{{\veci\da}}-1) (n_{{\vecj\ua}}-1)}+\nonumber \\ &&
\frac{(n_\veci-1) (n_\vecj-1) I_{{\veci\ua} \vecj}^{24}}{2 (n_{{\veci\ua}}-1) (n_{{\vecj\da}}-1)(n_{{\vecj\ua}}-1)}+
\frac{(n_\veci-1) (n_\vecj-1) I_{{\veci\ua} {\vecj\da}}^{22}}{4 (n_{{\veci\ua}}-1) (n_{{\vecj\da}}-1)}+
\frac{(n_\veci-1) (n_\vecj-1) I_{{\veci\ua} {\vecj\ua}}^{22}}{4 (n_{{\veci\ua}}-1) (n_{{\vecj\ua}}-1)}+\nonumber \\ &&
\frac{I_{{\vecj\ua}}^2 n_\veci (n_\vecj-1)}{4 (n_{{\vecj\ua}}-1)}+
\frac{I_{{\vecj\da}}^2 n_\veci (n_\vecj-1)}{4 (n_{{\vecj\da}}-1)}+
\frac{I_{{\veci\ua}}^2 (n_\veci-1) n_\vecj}{4 (n_{{\veci\ua}}-1)}+
\frac{I_\veci^4 (n_\veci-1) n_\vecj}{2 (n_{{\veci\da}}-1) (n_{{\veci\ua}}-1)}+\nonumber \\ &&
\frac{I_{{\veci \vecj}}^{44} (n_\veci-1) (n_\vecj-1)}{(n_{{\veci\da}}-1) (n_{{\veci\ua}}-1) (n_{{\vecj\da}}-1)(n_{{\vecj \ua}}-1)}+
\frac{I_{{\veci\da}}^2 (n_\veci-1) n_\vecj}{4 (n_{{\veci\da}}-1)}+
\frac{I_\vecj^4 n_\veci (n_\vecj-1)}{2 (n_{{\vecj\da}}-1) (n_{{\vecj\ua}}-1)}.
\eqnx
\eqn
\la \hat{S}_\veci^{z}\hat{S}_\vecj^{z} \ra_G &=& \frac{m_\veci m_\vecj}{4} +
\frac{m_\veci (m_\vecj+1) I_{\veci {\vecj\da}}^{42}}{2 (n_{{\veci\da}}-1) (n_{{\veci\ua}}-1)(n_{{\vecj\da}}-1)}-
   \frac{m_\veci (1-m_\vecj) I_{\veci {\vecj\ua}}^{42}}{2 (n_{{\veci\da}}-1) (n_{{\veci\ua}}-1) (n_{{\vecj\ua}}-1)}-\nonumber \\ &&
   \frac{(-m_\veci-1) m_\vecj I_{{\veci\da} \vecj}^{24}}{2 (n_{{\veci\da}}-1) (n_{{\vecj\da}}-1) (n_{{\vecj\ua}}-1)}+
   \frac{(-m_\veci-1) (-m_\vecj-1) I_{{\veci\da} {\vecj\da}}^{22}}{4 (n_{{\veci\da}}-1) (n_{{\vecj\da}}-1)}+
   \frac{(-m_\veci-1) (1-m_\vecj) I_{{\veci\da} {\vecj\ua}}^{22}}{4(n_{{\veci\da}}-1) (n_{{\vecj\ua}}-1)}-\nonumber \\ &&
   \frac{(-m_\veci+1) m_\vecj I_{{\veci\ua} \vecj}^{24}}{2(n_{{\veci\ua}}-1) (n_{{\vecj\da}}-1) (n_{{\vecj\ua}}-1)}+
   \frac{(-m_\veci+1)(-m_\vecj-1) I_{{\veci\ua} {\vecj\da}}^{22}}{4 (n_{{\veci\ua}}-1)(n_{{\vecj\da}}-1)}+
   \frac{(-m_\veci+1) (1-m_\vecj) I_{{\veci\ua} {\vecj\ua}}^{22}}{4 (n_{{\veci\ua}}-1) (n_{{\vecj\ua}}-1)}+\nonumber \\ &&
   \frac{I_{{\vecj\ua}}^2 m_\veci (1-m_\vecj)}{4(1-n_{{\vecj\ua}})}+
   \frac{I_{{\vecj\da}}^2 m_\veci (-m_\vecj-1)}{4(1-n_{{\vecj\da}})}-
   \frac{I_{{\veci\ua}}^2 (-m_\veci+1) m_\vecj}{4(n_{{\veci\ua}}-1)}+\nonumber \\ &&
   \frac{I_{\veci \vecj}^{44} m_\veci m_\vecj}{(n_{{\veci\da}}-1)(n_{{\veci\ua}}-1) (n_{{\vecj\da}}-1) (n_{{\vecj\ua}}-1)}-
   \frac{I_{{\veci\da}}^{2} (m_\veci+1) m_\vecj}{4 (1-n_{{\veci\da}})}+
   \frac{I^\vecj_4 m_\veci m_\vecj}{2 (1-n_{{\vecj\da}})(1-n_{{\vecj\ua}})}+\nonumber \\ &&
   \frac{I^\veci_4 m_\veci m_\vecj}{2 (1-n_{{\veci\da}})(1-n_{{\veci\ua}})}.
\eqnx

\section{Gutzwiller factors change} \label{app:GF}

\begin{figure}[ht!]
\centerline{\includegraphics[angle=270,width=0.8\columnwidth]{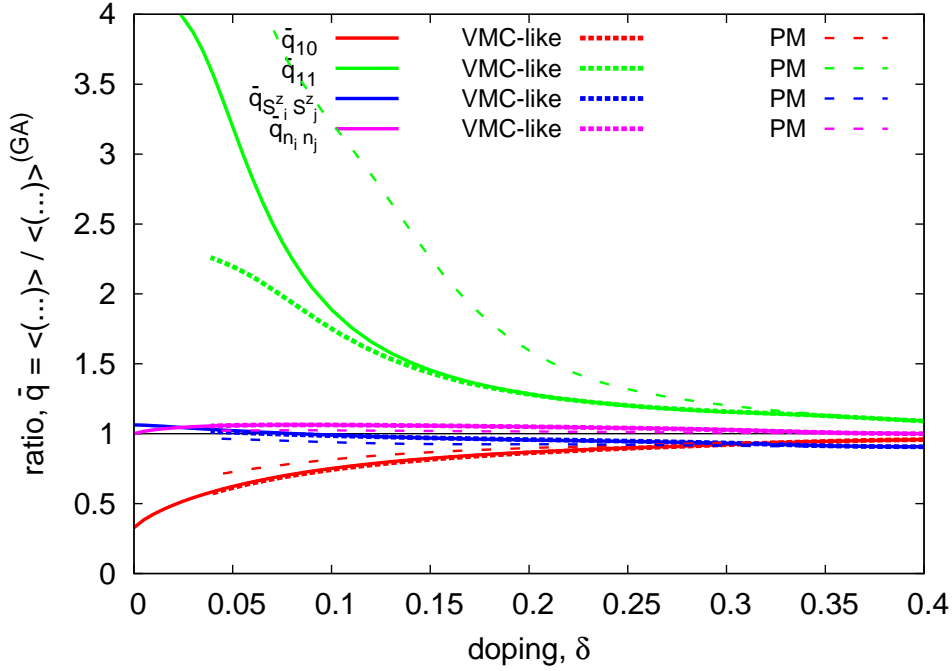}}
\caption{(Color online) The ratio of the averages $\la (...) \ra_G$ obtained in (VMC-like)DE-GWF with respect to that obtained in GA.
\label{fig:GF}}
\end{figure}
\tg{
In Fig. \ref{fig:GF} we plot the ratio of the averages $\la (...) \ra_G$ obtained accurately within (VMC-like)DE-GWF (Eqs.~(\ref{eq:iso1})-(\ref{eq:iso4})) and those obtained by within Gutzwiller approximation (Eqs.~(\ref{eq:iso1GA})-(\ref{eq:iso3GA})). Explicitly, we plot the following quantities
\eqn
\overline{q}_{\veci\vecj} &\equiv& \frac{\la \hat{c}^{\dagger}_{\veci,\sigma}\hat{c}_{\vecj,\sigma} \ra_G}{ \la \hat{c}^{\dagger}_{\veci,\sigma}\hat{c}_{\vecj,\sigma} \ra_G^{(GA)}} = \frac{q^2 T^{11}_{\veci\vecj} + 2 q \alpha T^{13}_{\veci\vecj} + \alpha^2 T^{33}_{\veci\vecj}}{q^2 P_{\veci\vecj}}, \\
\overline{q}_{S_i^z S_j^z} &\equiv& \frac{\la \hat{S}_\veci^{+}\hat{S}_\vecj^{-} + \hat{S}_\veci^{-}\hat{S}_\vecj^{+} \ra_G }{ \la \hat{S}_\veci^{+}\hat{S}_\vecj^{-} + \hat{S}_\veci^{-}\hat{S}_\vecj^{+} \ra_G^{(GA)}} = \frac{2 S^{22}}{\la \hat{S}_\veci^{+}\hat{S}_\vecj^{-} + \hat{S}_\veci^{-}\hat{S}_\vecj^{+} \ra_0} = \frac{S^{22}}{S^{22}(0)}, \\
\overline{q}_{n_i n_j} &\equiv& \frac{\la \hat{n}_\veci \hat{n}_\vecj \ra_G }{ \la \hat{n}_\veci \hat{n}_\vecj \ra_G^{(GA)}} = \frac{
n^2 + I^{22\ua\ua} \gamma + I^{22\ua\da} \gamma + (...)}{n^2 + I^{22\ua\ua}(0) \gamma + I^{22\ua\da}(0)\gamma}.
\eqnx
where $\gamma \equiv \frac{\left(1-2 n\right){}^2}{2 \left(n-1\right){}^2}$, by e.g. $S^{22}(0)$ we understand the zeroth-order diagrammatic sum, and by (...) we denote other diagrammatic sum terms, (see Eq. (\ref{eq:iso4})). According to the above expressions, a situation in which GA approximates the average accurately corresponds to $\overline{q} = 1$. If an average is overestimated (underestimated) by GA, this yields $\overline{q}<1$ ($\overline{q}>1$).
It can be seen from Fig.~\ref{fig:GF} that for the exchange term averages $\overline{q} \approx 1$, and therefore GA works quite well for them. However, for the kinetic energy term averages GA largely overestimates the n.n. average (as also reported in Ref. \onlinecite{Fukushima}) and underestimates the next n.n. average, especially for an underdoped system. This is the reason behind the large discrepancy of the GA and VMC results in this regime. The ratios $\overline{q}$ are quite similar in VMC-like and full DE-GWF methods. They are also similar in the PM phase (however, for the next n.n. hopping the ratio $\overline{q}_{11}$is substantially larger).
}

\section{Convergence analysis: number of lines} \label{app:1d}

\begin{figure}[ht!]
\centerline{\includegraphics[angle=270,width=0.8\columnwidth]{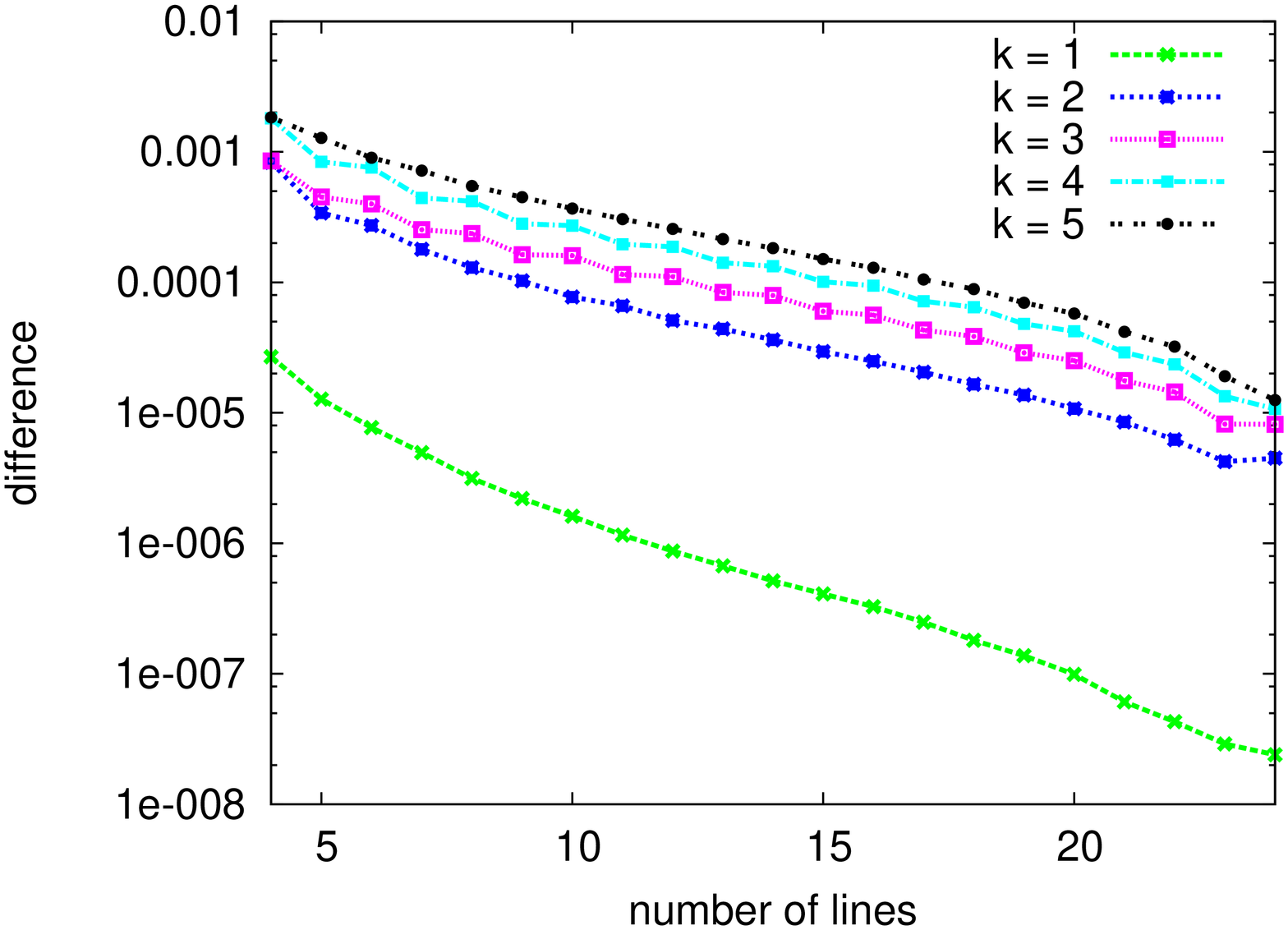}}
\caption{(Color online) Convergence of the results for $\la \hat{S}_\veci^z \hat{S}_\vecj^z \ra$ as a function of number of $\Ps$ lines.
\label{fig:2}}
\end{figure}

To analyze the effect of number of $\Ps$ lines included in the calculations we present in Fig. \ref{fig:2} the difference (integrated over doping values) between the correlation function $\la \hat{S}_\veci^z \hat{S}_\vecj^z \ra$ for a given number of lines $n$ and for 25 lines as a function of $n$.

Nearly linear behavior of the differences in Fig. \ref{fig:2} suggests that the convergence is exponential (a logarithmic scale is used in Fig. \ref{fig:2}). Note also that the higher-order results converge more slowly than the lower-order results, what indicates that to obtain the same accuracy (with respect to the complete $\Ps$ results with all lines included) in a higher order we need to take into account more lines than in a lower order. Therefore, not only the inclusion of higher-order terms is important to improve accuracy, but also the inclusion of longer range lines.

\section{Details of the VMC-like DE-GWF calculations} \label{app:B}

We set all parameters of the effective Hamiltonian to zero, except for n.n. pairing $\Delta^{\rm eff}_{10}$ and hoppings $t^{\rm eff}_{10}$, $t^{\rm eff}_{11}$, as well as $t^{\rm eff}_{00}$ playing the role of effective chemical potential. The n.n.~hopping is kept fixed, whereas the other parameters are optimized variationally. In the resulting scheme the effective Hamiltonian contains the same variational parameters as that used in VMC~\cite{PhysRevB.74.165109}.

\begin{figure}[ht!]
\centering
\includegraphics[width=0.54\columnwidth]{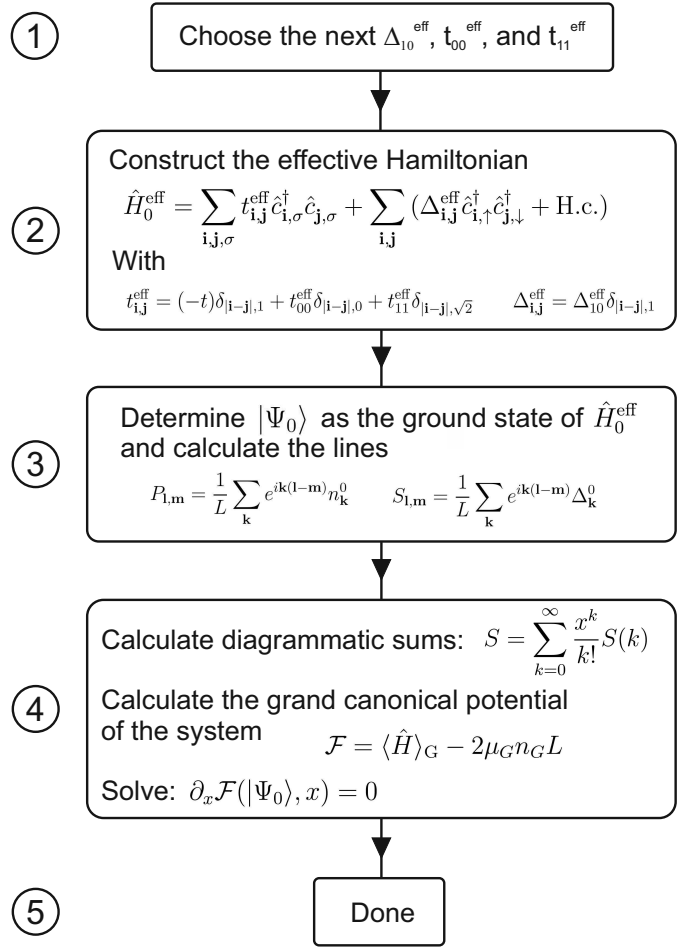}
\caption{Flowchart of the VMC-like DE-GWF calculations.}
\label{fig:S3}
\end{figure}

We have taken as nonzero the $|\Psi_0\rangle$ lines ($S_{\veci, \vecj} \equiv S_{0, (\veci - \vecj)} \equiv S_{XY}$ with $X = (i_1 - j_1), Y = (i_2 - j_2)$, $P_{XY}$ - analogously) fulfilling $X^2 + Y^2 \leq 25$. In the situation when the number of $|\Psi_0\rangle$ lines does not match the number of effective parameters ($t_{\veci,\vecj}^{\rm eff}$ and $\Delta_{\veci,\vecj}^{\rm eff}$), the self-consistency loop would not find the true minimum of the energy and a more standard minimization of the energy with respect to $\Delta^{\rm eff}_{10}$, $t^{\rm eff}_{00}$, and $t^{\rm eff}_{11}$ is necessary. Namely, we numerically search for a minimum of the system grand canonical potential $\mathcal{F}$ by calculating its value for fixed $\Delta^{\rm eff}_{10}$, $t^{\rm eff}_{00}$, and $t^{\rm eff}_{11}$. The flowchart of such calculations is presented in Fig. \ref{fig:S3}. Explicitly, having fixed effective parameters (step 1 in Fig.~\ref{fig:S3}) we may construct the effective Hamiltonian (step 2), calculate the $|\Psi_0\rangle$ lines (step 3), and having them we can obtain the diagrammatic sums and the potential $\mathcal{F}$ (step 4). Finally, we choose the solution with $\Delta^{\rm eff}_{10}$, $t^{\rm eff}_{00}$, and $t^{\rm eff}_{11}$ corresponding to the lowest potential $\mathcal{F}$.

\bibliography{DiagExp_tJ}

\providecommand{\noopsort}[1]{}\providecommand{\singleletter}[1]{#1}%
\begin{thebibliography}{79}%
\makeatletter
\providecommand \@ifxundefined [1]{%
 \@ifx{#1\undefined}
}%
\providecommand \@ifnum [1]{%
 \ifnum #1\expandafter \@firstoftwo
 \else \expandafter \@secondoftwo
 \fi
}%
\providecommand \@ifx [1]{%
 \ifx #1\expandafter \@firstoftwo
 \else \expandafter \@secondoftwo
 \fi
}%
\providecommand \natexlab [1]{#1}%
\providecommand \enquote  [1]{``#1''}%
\providecommand \bibnamefont  [1]{#1}%
\providecommand \bibfnamefont [1]{#1}%
\providecommand \citenamefont [1]{#1}%
\providecommand \href@noop [0]{\@secondoftwo}%
\providecommand \href [0]{\begingroup \@sanitize@url \@href}%
\providecommand \@href[1]{\@@startlink{#1}\@@href}%
\providecommand \@@href[1]{\endgroup#1\@@endlink}%
\providecommand \@sanitize@url [0]{\catcode `\\12\catcode `\$12\catcode
  `\&12\catcode `\#12\catcode `\^12\catcode `\_12\catcode `\%12\relax}%
\providecommand \@@startlink[1]{}%
\providecommand \@@endlink[0]{}%
\providecommand \url  [0]{\begingroup\@sanitize@url \@url }%
\providecommand \@url [1]{\endgroup\@href {#1}{\urlprefix }}%
\providecommand \urlprefix  [0]{URL }%
\providecommand \Eprint [0]{\href }%
\providecommand \doibase [0]{http://dx.doi.org/}%
\providecommand \selectlanguage [0]{\@gobble}%
\providecommand \bibinfo  [0]{\@secondoftwo}%
\providecommand \bibfield  [0]{\@secondoftwo}%
\providecommand \translation [1]{[#1]}%
\providecommand \BibitemOpen [0]{}%
\providecommand \bibitemStop [0]{}%
\providecommand \bibitemNoStop [0]{.\EOS\space}%
\providecommand \EOS [0]{\spacefactor3000\relax}%
\providecommand \BibitemShut  [1]{\csname bibitem#1\endcsname}%
\let\auto@bib@innerbib\@empty
\bibitem [{\citenamefont {Lee}\ \emph {et~al.}(2006)\citenamefont {Lee},
  \citenamefont {Nagaosa},\ and\ \citenamefont {Wen}}]{Rev_High_Tc}%
  \BibitemOpen
  \bibfield  {author} {\bibinfo {author} {\bibfnamefont {P.~A.}\ \bibnamefont
  {Lee}}, \bibinfo {author} {\bibfnamefont {N.}~\bibnamefont {Nagaosa}}, \ and\
  \bibinfo {author} {\bibfnamefont {X.-G.}\ \bibnamefont {Wen}},\ }\href
  {\doibase 10.1103/RevModPhys.78.17} {\bibfield  {journal} {\bibinfo
  {journal} {Rev. Mod. Phys.}\ }\textbf {\bibinfo {volume} {78}},\ \bibinfo
  {pages} {17} (\bibinfo {year} {2006})}\BibitemShut {NoStop}%
\bibitem [{\citenamefont {Ogata}\ and\ \citenamefont {Fukuyama}(2008)}]{Ogata}%
  \BibitemOpen
  \bibfield  {author} {\bibinfo {author} {\bibfnamefont {M.}~\bibnamefont
  {Ogata}}\ and\ \bibinfo {author} {\bibfnamefont {H.}~\bibnamefont
  {Fukuyama}},\ }\href@noop {} {\bibfield  {journal} {\bibinfo  {journal} {Rep.
  Prog. Phys.}\ }\textbf {\bibinfo {volume} {71}},\ \bibinfo {pages} {036501}
  (\bibinfo {year} {2008})}\BibitemShut {NoStop}%
\bibitem [{\citenamefont {Scalapino}(2012)}]{RevModPhys.84.1383}%
  \BibitemOpen
  \bibfield  {author} {\bibinfo {author} {\bibfnamefont {D.~J.}\ \bibnamefont
  {Scalapino}},\ }\href {\doibase 10.1103/RevModPhys.84.1383} {\bibfield
  {journal} {\bibinfo  {journal} {Rev. Mod. Phys.}\ }\textbf {\bibinfo {volume}
  {84}},\ \bibinfo {pages} {1383} (\bibinfo {year} {2012})}\BibitemShut
  {NoStop}%
\bibitem [{\citenamefont {Edegger}\ \emph {et~al.}(2007)\citenamefont
  {Edegger}, \citenamefont {Muthukumar},\ and\ \citenamefont
  {Gros}}]{EdeggerRev}%
  \BibitemOpen
  \bibfield  {author} {\bibinfo {author} {\bibfnamefont {B.}~\bibnamefont
  {Edegger}}, \bibinfo {author} {\bibfnamefont {V.~N.}\ \bibnamefont
  {Muthukumar}}, \ and\ \bibinfo {author} {\bibfnamefont {C.}~\bibnamefont
  {Gros}},\ }\href {\doibase 10.1080/00018730701627707} {\bibfield  {journal}
  {\bibinfo  {journal} {Adv. Phys.}\ }\textbf {\bibinfo {volume} {56}},\
  \bibinfo {pages} {927} (\bibinfo {year} {2007})}\BibitemShut {NoStop}%
\bibitem [{\citenamefont {Anderson}(2007)}]{Science.317.1704}%
  \BibitemOpen
  \bibfield  {author} {\bibinfo {author} {\bibfnamefont {P.}~\bibnamefont
  {Anderson}},\ }\href@noop {} {\bibfield  {journal} {\bibinfo  {journal}
  {Science}\ }\textbf {\bibinfo {volume} {317}},\ \bibinfo {pages} {1705}
  (\bibinfo {year} {2007})}\BibitemShut {NoStop}%
\bibitem [{\citenamefont {Spa{\l}ek}\ and\ \citenamefont
  {Goc-Jag{\l}o}(2012)}]{JSDGJ}%
  \BibitemOpen
  \bibfield  {author} {\bibinfo {author} {\bibfnamefont {J.}~\bibnamefont
  {Spa{\l}ek}}\ and\ \bibinfo {author} {\bibfnamefont {D.}~\bibnamefont
  {Goc-Jag{\l}o}},\ }\href@noop {} {\bibfield  {journal} {\bibinfo  {journal}
  {Phys. Scr.}\ }\textbf {\bibinfo {volume} {86}},\ \bibinfo {pages} {048301}
  (\bibinfo {year} {2012})}\BibitemShut {NoStop}%
\bibitem [{\citenamefont {Maier}\ \emph
  {et~al.}(2006{\natexlab{a}})\citenamefont {Maier}, \citenamefont {Jarrell},\
  and\ \citenamefont {Scalapino}}]{PhysRevLett.96.047005}%
  \BibitemOpen
  \bibfield  {author} {\bibinfo {author} {\bibfnamefont {T.~A.}\ \bibnamefont
  {Maier}}, \bibinfo {author} {\bibfnamefont {M.~S.}\ \bibnamefont {Jarrell}},
  \ and\ \bibinfo {author} {\bibfnamefont {D.~J.}\ \bibnamefont {Scalapino}},\
  }\href {\doibase 10.1103/PhysRevLett.96.047005} {\bibfield  {journal}
  {\bibinfo  {journal} {Phys. Rev. Lett.}\ }\textbf {\bibinfo {volume} {96}},\
  \bibinfo {pages} {047005} (\bibinfo {year} {2006}{\natexlab{a}})}\BibitemShut
  {NoStop}%
\bibitem [{\citenamefont {Maier}\ \emph
  {et~al.}(2006{\natexlab{b}})\citenamefont {Maier}, \citenamefont {Jarrell},\
  and\ \citenamefont {Scalapino}}]{PhysRevB.74.094513}%
  \BibitemOpen
  \bibfield  {author} {\bibinfo {author} {\bibfnamefont {T.~A.}\ \bibnamefont
  {Maier}}, \bibinfo {author} {\bibfnamefont {M.~S.}\ \bibnamefont {Jarrell}},
  \ and\ \bibinfo {author} {\bibfnamefont {D.~J.}\ \bibnamefont {Scalapino}},\
  }\href {\doibase 10.1103/PhysRevB.74.094513} {\bibfield  {journal} {\bibinfo
  {journal} {Phys. Rev. B}\ }\textbf {\bibinfo {volume} {74}},\ \bibinfo
  {pages} {094513} (\bibinfo {year} {2006}{\natexlab{b}})}\BibitemShut
  {NoStop}%
\bibitem [{\citenamefont {Monthoux}\ \emph {et~al.}(1991)\citenamefont
  {Monthoux}, \citenamefont {Balatsky},\ and\ \citenamefont
  {Pines}}]{PhysRevLett.67.3448}%
  \BibitemOpen
  \bibfield  {author} {\bibinfo {author} {\bibfnamefont {P.}~\bibnamefont
  {Monthoux}}, \bibinfo {author} {\bibfnamefont {A.~V.}\ \bibnamefont
  {Balatsky}}, \ and\ \bibinfo {author} {\bibfnamefont {D.}~\bibnamefont
  {Pines}},\ }\href {\doibase 10.1103/PhysRevLett.67.3448} {\bibfield
  {journal} {\bibinfo  {journal} {Phys. Rev. Lett.}\ }\textbf {\bibinfo
  {volume} {67}},\ \bibinfo {pages} {3448} (\bibinfo {year}
  {1991})}\BibitemShut {NoStop}%
\bibitem [{\citenamefont {Kyung}\ \emph {et~al.}(2009)\citenamefont {Kyung},
  \citenamefont {S\'en\'echal},\ and\ \citenamefont
  {Tremblay}}]{PhysRevB.80.205109}%
  \BibitemOpen
  \bibfield  {author} {\bibinfo {author} {\bibfnamefont {B.}~\bibnamefont
  {Kyung}}, \bibinfo {author} {\bibfnamefont {D.}~\bibnamefont {S\'en\'echal}},
  \ and\ \bibinfo {author} {\bibfnamefont {A.-M.~S.}\ \bibnamefont
  {Tremblay}},\ }\href {\doibase 10.1103/PhysRevB.80.205109} {\bibfield
  {journal} {\bibinfo  {journal} {Phys. Rev. B}\ }\textbf {\bibinfo {volume}
  {80}},\ \bibinfo {pages} {205109} (\bibinfo {year} {2009})}\BibitemShut
  {NoStop}%
\bibitem [{\citenamefont {Hanke}\ \emph {et~al.}(2010)\citenamefont {Hanke},
  \citenamefont {Kiesel}, \citenamefont {Aichhorn}, \citenamefont {Brehm},\
  and\ \citenamefont {Arrigoni}}]{Hanke1}%
  \BibitemOpen
  \bibfield  {author} {\bibinfo {author} {\bibfnamefont {W.}~\bibnamefont
  {Hanke}}, \bibinfo {author} {\bibfnamefont {M.}~\bibnamefont {Kiesel}},
  \bibinfo {author} {\bibfnamefont {M.}~\bibnamefont {Aichhorn}}, \bibinfo
  {author} {\bibfnamefont {S.}~\bibnamefont {Brehm}}, \ and\ \bibinfo {author}
  {\bibfnamefont {E.}~\bibnamefont {Arrigoni}},\ }\href {\doibase
  10.1140/epjst/e2010-01294-y} {\bibfield  {journal} {\bibinfo  {journal} {Eur.
  Phys. J. Special Topics}\ }\textbf {\bibinfo {volume} {188}},\ \bibinfo
  {pages} {15} (\bibinfo {year} {2010})}\BibitemShut {NoStop}%
\bibitem [{\citenamefont {Anderson}(1988)}]{AndersonBook}%
  \BibitemOpen
  \bibfield  {author} {\bibinfo {author} {\bibfnamefont {P.~W.}\ \bibnamefont
  {Anderson}},\ }\href@noop {} {\emph {\bibinfo {title} {Frontiers and
  Borderlines in Many-Particle Physics}}},\ edited by\ \bibinfo {editor}
  {\bibfnamefont {R.~A.}\ \bibnamefont {Broglia}}\ and\ \bibinfo {editor}
  {\bibfnamefont {J.~R.}\ \bibnamefont {Schrieffer}}\ (\bibinfo  {publisher}
  {North-Holland, Amsterdam},\ \bibinfo {year} {1988})\ pp.\ \bibinfo {pages}
  {1--40}\BibitemShut {NoStop}%
\bibitem [{\citenamefont {Zhang}\ and\ \citenamefont
  {Rice}(1988)}]{PhysRevB.37.3759}%
  \BibitemOpen
  \bibfield  {author} {\bibinfo {author} {\bibfnamefont {F.~C.}\ \bibnamefont
  {Zhang}}\ and\ \bibinfo {author} {\bibfnamefont {T.~M.}\ \bibnamefont
  {Rice}},\ }\href {\doibase 10.1103/PhysRevB.37.3759} {\bibfield  {journal}
  {\bibinfo  {journal} {Phys. Rev. B}\ }\textbf {\bibinfo {volume} {37}},\
  \bibinfo {pages} {3759} (\bibinfo {year} {1988})}\BibitemShut {NoStop}%
\bibitem [{\citenamefont {Zhang}\ and\ \citenamefont
  {Rice}(1990)}]{PhysRevB.41.7243}%
  \BibitemOpen
  \bibfield  {author} {\bibinfo {author} {\bibfnamefont {F.~C.}\ \bibnamefont
  {Zhang}}\ and\ \bibinfo {author} {\bibfnamefont {T.~M.}\ \bibnamefont
  {Rice}},\ }\href {\doibase 10.1103/PhysRevB.41.7243} {\bibfield  {journal}
  {\bibinfo  {journal} {Phys. Rev. B}\ }\textbf {\bibinfo {volume} {41}},\
  \bibinfo {pages} {7243} (\bibinfo {year} {1990})}\BibitemShut {NoStop}%
\bibitem [{\citenamefont {Spa\l{}ek}(1988{\natexlab{a}})}]{Spalek1}%
  \BibitemOpen
  \bibfield  {author} {\bibinfo {author} {\bibfnamefont {J.}~\bibnamefont
  {Spa\l{}ek}},\ }\href {\doibase 10.1103/PhysRevB.37.533} {\bibfield
  {journal} {\bibinfo  {journal} {Phys. Rev. B}\ }\textbf {\bibinfo {volume}
  {37}},\ \bibinfo {pages} {533} (\bibinfo {year}
  {1988}{\natexlab{a}})}\BibitemShut {NoStop}%
\bibitem [{\citenamefont {Spa\l{}ek}(1988{\natexlab{b}})}]{Spalek2}%
  \BibitemOpen
  \bibfield  {author} {\bibinfo {author} {\bibfnamefont {J.}~\bibnamefont
  {Spa\l{}ek}},\ }\href {\doibase 10.1103/PhysRevB.38.208} {\bibfield
  {journal} {\bibinfo  {journal} {Phys. Rev. B}\ }\textbf {\bibinfo {volume}
  {38}},\ \bibinfo {pages} {208} (\bibinfo {year}
  {1988}{\natexlab{b}})}\BibitemShut {NoStop}%
\bibitem [{\citenamefont {Gutzwiller}(1963)}]{Gutzwiller}%
  \BibitemOpen
  \bibfield  {author} {\bibinfo {author} {\bibfnamefont {M.~C.}\ \bibnamefont
  {Gutzwiller}},\ }\href {\doibase 10.1103/PhysRevLett.10.159} {\bibfield
  {journal} {\bibinfo  {journal} {Phys. Rev. Lett.}\ }\textbf {\bibinfo
  {volume} {10}},\ \bibinfo {pages} {159} (\bibinfo {year} {1963})}\BibitemShut
  {NoStop}%
\bibitem [{\citenamefont {Gutzwiller}(1965)}]{PhysRev.137.A1726}%
  \BibitemOpen
  \bibfield  {author} {\bibinfo {author} {\bibfnamefont {M.~C.}\ \bibnamefont
  {Gutzwiller}},\ }\href {\doibase 10.1103/PhysRev.137.A1726} {\bibfield
  {journal} {\bibinfo  {journal} {Phys. Rev.}\ }\textbf {\bibinfo {volume}
  {137}},\ \bibinfo {pages} {A1726} (\bibinfo {year} {1965})}\BibitemShut
  {NoStop}%
\bibitem [{\citenamefont {Zhang}\ \emph {et~al.}(1988)\citenamefont {Zhang},
  \citenamefont {Gros}, \citenamefont {Rice},\ and\ \citenamefont
  {Shiba}}]{Zhang}%
  \BibitemOpen
  \bibfield  {author} {\bibinfo {author} {\bibfnamefont {F.~C.}\ \bibnamefont
  {Zhang}}, \bibinfo {author} {\bibfnamefont {C.}~\bibnamefont {Gros}},
  \bibinfo {author} {\bibfnamefont {T.~M.}\ \bibnamefont {Rice}}, \ and\
  \bibinfo {author} {\bibfnamefont {H.}~\bibnamefont {Shiba}},\ }\href
  {\doibase 10.1088/0953-2048/1/1/009} {\bibfield  {journal} {\bibinfo
  {journal} {Supercond. Sci. Technol.}\ }\textbf {\bibinfo {volume} {1}},\
  \bibinfo {pages} {36} (\bibinfo {year} {1988})}\BibitemShut {NoStop}%
\bibitem [{\citenamefont {Ogata}\ and\ \citenamefont {Himeda}(2003)}]{Ogata2}%
  \BibitemOpen
  \bibfield  {author} {\bibinfo {author} {\bibfnamefont {M.}~\bibnamefont
  {Ogata}}\ and\ \bibinfo {author} {\bibfnamefont {A.}~\bibnamefont {Himeda}},\
  }\href {\doibase 10.1143/JPSJ.72.374} {\bibfield  {journal} {\bibinfo
  {journal} {J. Phys. Soc. Jpn.}\ }\textbf {\bibinfo {volume} {72}},\ \bibinfo
  {pages} {374} (\bibinfo {year} {2003})}\BibitemShut {NoStop}%
\bibitem [{\citenamefont {Wang}\ \emph {et~al.}(2006)\citenamefont {Wang},
  \citenamefont {Wang}, \citenamefont {Chen},\ and\ \citenamefont
  {Zhang}}]{Wang}%
  \BibitemOpen
  \bibfield  {author} {\bibinfo {author} {\bibfnamefont {Q.-H.}\ \bibnamefont
  {Wang}}, \bibinfo {author} {\bibfnamefont {Z.~D.}\ \bibnamefont {Wang}},
  \bibinfo {author} {\bibfnamefont {Y.}~\bibnamefont {Chen}}, \ and\ \bibinfo
  {author} {\bibfnamefont {F.~C.}\ \bibnamefont {Zhang}},\ }\href {\doibase
  10.1103/PhysRevB.73.092507} {\bibfield  {journal} {\bibinfo  {journal} {Phys.
  Rev. B}\ }\textbf {\bibinfo {volume} {73}},\ \bibinfo {pages} {092507}
  (\bibinfo {year} {2006})}\BibitemShut {NoStop}%
\bibitem [{\citenamefont {Fukushima}(2008)}]{Fukushima}%
  \BibitemOpen
  \bibfield  {author} {\bibinfo {author} {\bibfnamefont {N.}~\bibnamefont
  {Fukushima}},\ }\href {\doibase 10.1103/PhysRevB.78.115105} {\bibfield
  {journal} {\bibinfo  {journal} {Phys. Rev. B}\ }\textbf {\bibinfo {volume}
  {78}},\ \bibinfo {pages} {115105} (\bibinfo {year} {2008})}\BibitemShut
  {NoStop}%
\bibitem [{\citenamefont {Fukushima}(2011)}]{Fukushima2}%
  \BibitemOpen
  \bibfield  {author} {\bibinfo {author} {\bibfnamefont {N.}~\bibnamefont
  {Fukushima}},\ }\href {\doibase 10.1088/1751-8113/44/7/075002} {\bibfield
  {journal} {\bibinfo  {journal} {J. Phys. A: Math. Theor.}\ }\textbf {\bibinfo
  {volume} {44}},\ \bibinfo {pages} {075002} (\bibinfo {year}
  {2011})}\BibitemShut {NoStop}%
\bibitem [{\citenamefont {J\c{e}drak}\ and\ \citenamefont
  {Spa{\l}ek}(2011)}]{Jedrak2}%
  \BibitemOpen
  \bibfield  {author} {\bibinfo {author} {\bibfnamefont {J.}~\bibnamefont
  {J\c{e}drak}}\ and\ \bibinfo {author} {\bibfnamefont {J.}~\bibnamefont
  {Spa{\l}ek}},\ }\href {\doibase 10.1103/PhysRevB.83.104512} {\bibfield
  {journal} {\bibinfo  {journal} {Phys. Rev. B}\ }\textbf {\bibinfo {volume}
  {83}},\ \bibinfo {pages} {104512} (\bibinfo {year} {2011})}\BibitemShut
  {NoStop}%
\bibitem [{\citenamefont {J\c{e}drak}\ and\ \citenamefont
  {Spa{\l}ek}(2010)}]{Jedrak1}%
  \BibitemOpen
  \bibfield  {author} {\bibinfo {author} {\bibfnamefont {J.}~\bibnamefont
  {J\c{e}drak}}\ and\ \bibinfo {author} {\bibfnamefont {J.}~\bibnamefont
  {Spa{\l}ek}},\ }\href {\doibase 10.1103/PhysRevB.81.073108} {\bibfield
  {journal} {\bibinfo  {journal} {Phys. Rev. B}\ }\textbf {\bibinfo {volume}
  {81}},\ \bibinfo {pages} {073108} (\bibinfo {year} {2010})}\BibitemShut
  {NoStop}%
\bibitem [{\citenamefont {Kaczmarczyk}\ and\ \citenamefont
  {Spa\l{}ek}(2011)}]{PhysRevB.84.125140}%
  \BibitemOpen
  \bibfield  {author} {\bibinfo {author} {\bibfnamefont {J.}~\bibnamefont
  {Kaczmarczyk}}\ and\ \bibinfo {author} {\bibfnamefont {J.}~\bibnamefont
  {Spa\l{}ek}},\ }\href {\doibase 10.1103/PhysRevB.84.125140} {\bibfield
  {journal} {\bibinfo  {journal} {Phys. Rev. B}\ }\textbf {\bibinfo {volume}
  {84}},\ \bibinfo {pages} {125140} (\bibinfo {year} {2011})}\BibitemShut
  {NoStop}%
\bibitem [{\citenamefont {Howczak}\ \emph {et~al.}(2013)\citenamefont
  {Howczak}, \citenamefont {Kaczmarczyk},\ and\ \citenamefont
  {Spa{\l}ek}}]{Olga}%
  \BibitemOpen
  \bibfield  {author} {\bibinfo {author} {\bibfnamefont {O.}~\bibnamefont
  {Howczak}}, \bibinfo {author} {\bibfnamefont {J.}~\bibnamefont
  {Kaczmarczyk}}, \ and\ \bibinfo {author} {\bibfnamefont {J.}~\bibnamefont
  {Spa{\l}ek}},\ }\href {\doibase 10.1002/pssb.201200774} {\bibfield  {journal}
  {\bibinfo  {journal} {Phys. Status Solidi (b)}\ }\textbf {\bibinfo {volume}
  {250}},\ \bibinfo {pages} {609} (\bibinfo {year} {2013})}\BibitemShut
  {NoStop}%
\bibitem [{\citenamefont {Zegrodnik}\ \emph {et~al.}(2013)\citenamefont
  {Zegrodnik}, \citenamefont {Spa{\l}ek},\ and\ \citenamefont
  {B\"{u}nemann}}]{Zegrodnik}%
  \BibitemOpen
  \bibfield  {author} {\bibinfo {author} {\bibfnamefont {M.}~\bibnamefont
  {Zegrodnik}}, \bibinfo {author} {\bibfnamefont {J.}~\bibnamefont
  {Spa{\l}ek}}, \ and\ \bibinfo {author} {\bibfnamefont {J.}~\bibnamefont
  {B\"{u}nemann}},\ }\href {\doibase 10.1088/1367-2630/15/7/073050} {\bibfield
  {journal} {\bibinfo  {journal} {New J. Phys.}\ }\textbf {\bibinfo {volume}
  {15}},\ \bibinfo {pages} {073050} (\bibinfo {year} {2013})}\BibitemShut
  {NoStop}%
\bibitem [{\citenamefont {Abram}\ \emph {et~al.}(2013)\citenamefont {Abram},
  \citenamefont {Kaczmarczyk}, \citenamefont {J\c{e}drak},\ and\ \citenamefont
  {Spa\l{}ek}}]{Abram}%
  \BibitemOpen
  \bibfield  {author} {\bibinfo {author} {\bibfnamefont {M.}~\bibnamefont
  {Abram}}, \bibinfo {author} {\bibfnamefont {J.}~\bibnamefont {Kaczmarczyk}},
  \bibinfo {author} {\bibfnamefont {J.}~\bibnamefont {J\c{e}drak}}, \ and\
  \bibinfo {author} {\bibfnamefont {J.}~\bibnamefont {Spa\l{}ek}},\ }\href
  {\doibase 10.1103/PhysRevB.88.094502} {\bibfield  {journal} {\bibinfo
  {journal} {Phys. Rev. B}\ }\textbf {\bibinfo {volume} {88}},\ \bibinfo
  {pages} {094502} (\bibinfo {year} {2013})}\BibitemShut {NoStop}%
\bibitem [{\citenamefont {B\"{u}nemann}\ \emph {et~al.}(2012)\citenamefont
  {B\"{u}nemann}, \citenamefont {Schickling},\ and\ \citenamefont
  {Gebhard}}]{Buenemann}%
  \BibitemOpen
  \bibfield  {author} {\bibinfo {author} {\bibfnamefont {J.}~\bibnamefont
  {B\"{u}nemann}}, \bibinfo {author} {\bibfnamefont {T.}~\bibnamefont
  {Schickling}}, \ and\ \bibinfo {author} {\bibfnamefont {F.}~\bibnamefont
  {Gebhard}},\ }\href@noop {} {\bibfield  {journal} {\bibinfo  {journal}
  {Europhys. Lett.}\ }\textbf {\bibinfo {volume} {98}},\ \bibinfo {pages}
  {27006} (\bibinfo {year} {2012})}\BibitemShut {NoStop}%
\bibitem [{\citenamefont {Kaczmarczyk}\ \emph {et~al.}(2013)\citenamefont
  {Kaczmarczyk}, \citenamefont {Spa\l{}ek}, \citenamefont {Schickling},\ and\
  \citenamefont {B\"unemann}}]{PhysRevB.88.115127}%
  \BibitemOpen
  \bibfield  {author} {\bibinfo {author} {\bibfnamefont {J.}~\bibnamefont
  {Kaczmarczyk}}, \bibinfo {author} {\bibfnamefont {J.}~\bibnamefont
  {Spa\l{}ek}}, \bibinfo {author} {\bibfnamefont {T.}~\bibnamefont
  {Schickling}}, \ and\ \bibinfo {author} {\bibfnamefont {J.}~\bibnamefont
  {B\"unemann}},\ }\href {\doibase 10.1103/PhysRevB.88.115127} {\bibfield
  {journal} {\bibinfo  {journal} {Phys. Rev. B}\ }\textbf {\bibinfo {volume}
  {88}},\ \bibinfo {pages} {115127} (\bibinfo {year} {2013})}\BibitemShut
  {NoStop}%
\bibitem [{\citenamefont {Metzner}\ and\ \citenamefont
  {Vollhardt}(1988)}]{Metzner}%
  \BibitemOpen
  \bibfield  {author} {\bibinfo {author} {\bibfnamefont {W.}~\bibnamefont
  {Metzner}}\ and\ \bibinfo {author} {\bibfnamefont {D.}~\bibnamefont
  {Vollhardt}},\ }\href {\doibase 10.1103/PhysRevB.37.7382} {\bibfield
  {journal} {\bibinfo  {journal} {Phys. Rev. B}\ }\textbf {\bibinfo {volume}
  {37}},\ \bibinfo {pages} {7382} (\bibinfo {year} {1988})}\BibitemShut
  {NoStop}%
\bibitem [{\citenamefont {Kurzyk}\ \emph {et~al.}(2007)\citenamefont {Kurzyk},
  \citenamefont {Spa\l{}ek},\ and\ \citenamefont {W\'{o}jcik}}]{Kurzyk}%
  \BibitemOpen
  \bibfield  {author} {\bibinfo {author} {\bibfnamefont {J.}~\bibnamefont
  {Kurzyk}}, \bibinfo {author} {\bibfnamefont {J.}~\bibnamefont {Spa\l{}ek}}, \
  and\ \bibinfo {author} {\bibfnamefont {J.}~\bibnamefont {W\'{o}jcik}},\
  }\href@noop {} {\bibfield  {journal} {\bibinfo  {journal} {Acta Phys. Polon.
  A}\ }\textbf {\bibinfo {volume} {111}},\ \bibinfo {pages} {603} (\bibinfo
  {year} {2007})}\BibitemShut {NoStop}%
\bibitem [{Note1()}]{Note1}%
  \BibitemOpen
  \bibinfo {note} {For original derivation of the $t$-$J$ model from the
  Hubbard model see: K. A. Chao, J. Spa{\l }ek, and A. M. Ole\'{s}, J. Phys. C
  \protect \textbf {10}, L271 (1977). For didactical exposition see: J. Spa{\l
  }ek, Acta Phys. Polon. A \protect \textbf {121}, 764 (2012).}\BibitemShut
  {Stop}%
\bibitem [{\citenamefont {Gebhard}(1990)}]{Gebhard}%
  \BibitemOpen
  \bibfield  {author} {\bibinfo {author} {\bibfnamefont {F.}~\bibnamefont
  {Gebhard}},\ }\href@noop {} {\bibfield  {journal} {\bibinfo  {journal} {Phys.
  Rev. B}\ }\textbf {\bibinfo {volume} {41}},\ \bibinfo {pages} {9452}
  (\bibinfo {year} {1990})}\BibitemShut {NoStop}%
\bibitem [{\citenamefont {Fetter}\ and\ \citenamefont
  {Walecka}(2003)}]{Fetter}%
  \BibitemOpen
  \bibfield  {author} {\bibinfo {author} {\bibfnamefont {A.~L.}\ \bibnamefont
  {Fetter}}\ and\ \bibinfo {author} {\bibfnamefont {J.~D.}\ \bibnamefont
  {Walecka}},\ }\href@noop {} {\emph {\bibinfo {title} {Quantum Theory of
  Many-Particle Systems}}}\ (\bibinfo  {publisher} {Dover Publications, New
  York},\ \bibinfo {year} {2003})\BibitemShut {NoStop}%
\bibitem [{Note2()}]{Note2}%
  \BibitemOpen
  \bibinfo {note} {{The numerical difference between the results of the zeroth
  order DE-GWF and GCGA is smaller than the line thickness of the presented
  curves, so neglecting of the diagrammatic sum $I_{{{\protect \bf i}{\protect
  \bf j}}}^{44}$ is not essential.}}\BibitemShut {Stop}%
\bibitem [{\citenamefont {Gebhard}\ and\ \citenamefont
  {Vollhardt}(1988)}]{PhysRevB.38.6911}%
  \BibitemOpen
  \bibfield  {author} {\bibinfo {author} {\bibfnamefont {F.}~\bibnamefont
  {Gebhard}}\ and\ \bibinfo {author} {\bibfnamefont {D.}~\bibnamefont
  {Vollhardt}},\ }\href {\doibase 10.1103/PhysRevB.38.6911} {\bibfield
  {journal} {\bibinfo  {journal} {Phys. Rev. B}\ }\textbf {\bibinfo {volume}
  {38}},\ \bibinfo {pages} {6911} (\bibinfo {year} {1988})}\BibitemShut
  {NoStop}%
\bibitem [{\citenamefont {Yang}\ \emph {et~al.}(2009)\citenamefont {Yang},
  \citenamefont {Chen}, \citenamefont {Rice}, \citenamefont {Sigrist},\ and\
  \citenamefont {Zhang}}]{Yang}%
  \BibitemOpen
  \bibfield  {author} {\bibinfo {author} {\bibfnamefont {K.-Y.}\ \bibnamefont
  {Yang}}, \bibinfo {author} {\bibfnamefont {W.~Q.}\ \bibnamefont {Chen}},
  \bibinfo {author} {\bibfnamefont {T.~M.}\ \bibnamefont {Rice}}, \bibinfo
  {author} {\bibfnamefont {M.}~\bibnamefont {Sigrist}}, \ and\ \bibinfo
  {author} {\bibfnamefont {F.-C.}\ \bibnamefont {Zhang}},\ }\href {\doibase
  10.1088/1367-2630/11/5/055053} {\bibfield  {journal} {\bibinfo  {journal}
  {New J. Phys.}\ }\textbf {\bibinfo {volume} {11}},\ \bibinfo {pages} {055053}
  (\bibinfo {year} {2009})}\BibitemShut {NoStop}%
\bibitem [{\citenamefont {B\"unemann}\ \emph {et~al.}(2003)\citenamefont
  {B\"unemann}, \citenamefont {Gebhard},\ and\ \citenamefont
  {Thul}}]{PhysRevB.67.075103}%
  \BibitemOpen
  \bibfield  {author} {\bibinfo {author} {\bibfnamefont {J.}~\bibnamefont
  {B\"unemann}}, \bibinfo {author} {\bibfnamefont {F.}~\bibnamefont {Gebhard}},
  \ and\ \bibinfo {author} {\bibfnamefont {R.}~\bibnamefont {Thul}},\ }\href
  {\doibase 10.1103/PhysRevB.67.075103} {\bibfield  {journal} {\bibinfo
  {journal} {Phys. Rev. B}\ }\textbf {\bibinfo {volume} {67}},\ \bibinfo
  {pages} {075103} (\bibinfo {year} {2003})}\BibitemShut {NoStop}%
\bibitem [{\citenamefont {Prokof'ev}\ and\ \citenamefont
  {Svistunov}(2008)}]{PhysRevB.77.125101}%
  \BibitemOpen
  \bibfield  {author} {\bibinfo {author} {\bibfnamefont {N.~V.}\ \bibnamefont
  {Prokof'ev}}\ and\ \bibinfo {author} {\bibfnamefont {B.~V.}\ \bibnamefont
  {Svistunov}},\ }\href {\doibase 10.1103/PhysRevB.77.125101} {\bibfield
  {journal} {\bibinfo  {journal} {Phys. Rev. B}\ }\textbf {\bibinfo {volume}
  {77}},\ \bibinfo {pages} {125101} (\bibinfo {year} {2008})}\BibitemShut
  {NoStop}%
\bibitem [{\citenamefont {Van~Houcke}\ \emph {et~al.}(2012)\citenamefont
  {Van~Houcke}, \citenamefont {Werner}, \citenamefont {Kozik}, \citenamefont
  {Prokof'ev}, \citenamefont {Svistunov}, \citenamefont {Ku}, \citenamefont
  {Sommer}, \citenamefont {Cheuk}, \citenamefont {Schirotzek},\ and\
  \citenamefont {Zwierlein}}]{Houcke}%
  \BibitemOpen
  \bibfield  {author} {\bibinfo {author} {\bibfnamefont {K.}~\bibnamefont
  {Van~Houcke}}, \bibinfo {author} {\bibfnamefont {F.}~\bibnamefont {Werner}},
  \bibinfo {author} {\bibfnamefont {E.}~\bibnamefont {Kozik}}, \bibinfo
  {author} {\bibfnamefont {N.}~\bibnamefont {Prokof'ev}}, \bibinfo {author}
  {\bibfnamefont {B.}~\bibnamefont {Svistunov}}, \bibinfo {author}
  {\bibfnamefont {M.~J.~H.}\ \bibnamefont {Ku}}, \bibinfo {author}
  {\bibfnamefont {A.~T.}\ \bibnamefont {Sommer}}, \bibinfo {author}
  {\bibfnamefont {L.~W.}\ \bibnamefont {Cheuk}}, \bibinfo {author}
  {\bibfnamefont {A.}~\bibnamefont {Schirotzek}}, \ and\ \bibinfo {author}
  {\bibfnamefont {M.~W.}\ \bibnamefont {Zwierlein}},\ }\href {\doibase
  10.1038/nphys2273} {\bibfield  {journal} {\bibinfo  {journal} {Nature Phys.}\
  }\textbf {\bibinfo {volume} {8}},\ \bibinfo {pages} {366} (\bibinfo {year}
  {2012})}\BibitemShut {NoStop}%
\bibitem [{\citenamefont {Edegger}\ \emph
  {et~al.}(2006{\natexlab{a}})\citenamefont {Edegger}, \citenamefont
  {Muthukumar},\ and\ \citenamefont {Gros}}]{PhysRevB.74.165109}%
  \BibitemOpen
  \bibfield  {author} {\bibinfo {author} {\bibfnamefont {B.}~\bibnamefont
  {Edegger}}, \bibinfo {author} {\bibfnamefont {V.~N.}\ \bibnamefont
  {Muthukumar}}, \ and\ \bibinfo {author} {\bibfnamefont {C.}~\bibnamefont
  {Gros}},\ }\href {\doibase 10.1103/PhysRevB.74.165109} {\bibfield  {journal}
  {\bibinfo  {journal} {Phys. Rev. B}\ }\textbf {\bibinfo {volume} {74}},\
  \bibinfo {pages} {165109} (\bibinfo {year} {2006}{\natexlab{a}})}\BibitemShut
  {NoStop}%
\bibitem [{\citenamefont {Watanabe}\ \emph {et~al.}(2009)\citenamefont
  {Watanabe}, \citenamefont {Yokoyama}, \citenamefont {Shigeta},\ and\
  \citenamefont {Ogata}}]{Watanabe}%
  \BibitemOpen
  \bibfield  {author} {\bibinfo {author} {\bibfnamefont {T.}~\bibnamefont
  {Watanabe}}, \bibinfo {author} {\bibfnamefont {H.}~\bibnamefont {Yokoyama}},
  \bibinfo {author} {\bibfnamefont {K.}~\bibnamefont {Shigeta}}, \ and\
  \bibinfo {author} {\bibfnamefont {M.}~\bibnamefont {Ogata}},\ }\href
  {\doibase 10.1088/1367-2630/11/7/075011} {\bibfield  {journal} {\bibinfo
  {journal} {New J. Phys.}\ }\textbf {\bibinfo {volume} {11}},\ \bibinfo
  {pages} {075011} (\bibinfo {year} {2009})}\BibitemShut {NoStop}%
\bibitem [{\citenamefont {Ivanov}(2004)}]{PhysRevB.70.104503}%
  \BibitemOpen
  \bibfield  {author} {\bibinfo {author} {\bibfnamefont {D.~A.}\ \bibnamefont
  {Ivanov}},\ }\href {\doibase 10.1103/PhysRevB.70.104503} {\bibfield
  {journal} {\bibinfo  {journal} {Phys. Rev. B}\ }\textbf {\bibinfo {volume}
  {70}},\ \bibinfo {pages} {104503} (\bibinfo {year} {2004})}\BibitemShut
  {NoStop}%
\bibitem [{\citenamefont {Raczkowski}\ \emph {et~al.}(2007)\citenamefont
  {Raczkowski}, \citenamefont {Capello}, \citenamefont {Poilblanc},
  \citenamefont {Fr\'esard},\ and\ \citenamefont {Ole\ifmmode~\acute{s}\else
  \'{s}\fi{}}}]{PhysRevB.76.140505}%
  \BibitemOpen
  \bibfield  {author} {\bibinfo {author} {\bibfnamefont {M.}~\bibnamefont
  {Raczkowski}}, \bibinfo {author} {\bibfnamefont {M.}~\bibnamefont {Capello}},
  \bibinfo {author} {\bibfnamefont {D.}~\bibnamefont {Poilblanc}}, \bibinfo
  {author} {\bibfnamefont {R.}~\bibnamefont {Fr\'esard}}, \ and\ \bibinfo
  {author} {\bibfnamefont {A.~M.}\ \bibnamefont {Ole\ifmmode~\acute{s}\else
  \'{s}\fi{}}},\ }\href {\doibase 10.1103/PhysRevB.76.140505} {\bibfield
  {journal} {\bibinfo  {journal} {Phys. Rev. B}\ }\textbf {\bibinfo {volume}
  {76}},\ \bibinfo {pages} {140505} (\bibinfo {year} {2007})}\BibitemShut
  {NoStop}%
\bibitem [{\citenamefont {Chou}\ and\ \citenamefont
  {Lee}(2012)}]{PhysRevB.85.104511}%
  \BibitemOpen
  \bibfield  {author} {\bibinfo {author} {\bibfnamefont {C.-P.}\ \bibnamefont
  {Chou}}\ and\ \bibinfo {author} {\bibfnamefont {T.-K.}\ \bibnamefont {Lee}},\
  }\href {\doibase 10.1103/PhysRevB.85.104511} {\bibfield  {journal} {\bibinfo
  {journal} {Phys. Rev. B}\ }\textbf {\bibinfo {volume} {85}},\ \bibinfo
  {pages} {104511} (\bibinfo {year} {2012})}\BibitemShut {NoStop}%
\bibitem [{\citenamefont {Edegger}\ \emph
  {et~al.}(2006{\natexlab{b}})\citenamefont {Edegger}, \citenamefont
  {Muthukumar}, \citenamefont {Gros},\ and\ \citenamefont
  {Anderson}}]{EdeggerPRL}%
  \BibitemOpen
  \bibfield  {author} {\bibinfo {author} {\bibfnamefont {B.}~\bibnamefont
  {Edegger}}, \bibinfo {author} {\bibfnamefont {V.~N.}\ \bibnamefont
  {Muthukumar}}, \bibinfo {author} {\bibfnamefont {C.}~\bibnamefont {Gros}}, \
  and\ \bibinfo {author} {\bibfnamefont {P.~W.}\ \bibnamefont {Anderson}},\
  }\href {\doibase 10.1103/PhysRevLett.96.207002} {\bibfield  {journal}
  {\bibinfo  {journal} {Phys. Rev. Lett.}\ }\textbf {\bibinfo {volume} {96}},\
  \bibinfo {pages} {207002} (\bibinfo {year} {2006}{\natexlab{b}})}\BibitemShut
  {NoStop}%
\bibitem [{\citenamefont {Yunoki}\ \emph {et~al.}(2005)\citenamefont {Yunoki},
  \citenamefont {Dagotto},\ and\ \citenamefont {Sorella}}]{Yunoki}%
  \BibitemOpen
  \bibfield  {author} {\bibinfo {author} {\bibfnamefont {S.}~\bibnamefont
  {Yunoki}}, \bibinfo {author} {\bibfnamefont {E.}~\bibnamefont {Dagotto}}, \
  and\ \bibinfo {author} {\bibfnamefont {S.}~\bibnamefont {Sorella}},\ }\href
  {\doibase 10.1103/PhysRevLett.94.037001} {\bibfield  {journal} {\bibinfo
  {journal} {Phys. Rev. Lett.}\ }\textbf {\bibinfo {volume} {94}},\ \bibinfo
  {pages} {037001} (\bibinfo {year} {2005})}\BibitemShut {NoStop}%
\bibitem [{\citenamefont {Zhou}\ \emph {et~al.}(2003)\citenamefont {Zhou},
  \citenamefont {Yoshida}, \citenamefont {Lanzara}, \citenamefont {Bogdanov},
  \citenamefont {Kellar}, \citenamefont {Shen}, \citenamefont {Yang},
  \citenamefont {Ronning}, \citenamefont {Sasagawa}, \citenamefont {Kakeshita},
  \citenamefont {Noda}, \citenamefont {Eisaki}, \citenamefont {Uchida},
  \citenamefont {Lin}, \citenamefont {Zhou}, \citenamefont {Xiong},
  \citenamefont {Ti}, \citenamefont {Zhao}, \citenamefont {Fujimori},
  \citenamefont {Hussain},\ and\ \citenamefont {Shen}}]{Zhou}%
  \BibitemOpen
  \bibfield  {author} {\bibinfo {author} {\bibfnamefont {X.~J.}\ \bibnamefont
  {Zhou}}, \bibinfo {author} {\bibfnamefont {T.}~\bibnamefont {Yoshida}},
  \bibinfo {author} {\bibfnamefont {A.}~\bibnamefont {Lanzara}}, \bibinfo
  {author} {\bibfnamefont {P.~V.}\ \bibnamefont {Bogdanov}}, \bibinfo {author}
  {\bibfnamefont {S.~A.}\ \bibnamefont {Kellar}}, \bibinfo {author}
  {\bibfnamefont {K.~M.}\ \bibnamefont {Shen}}, \bibinfo {author}
  {\bibfnamefont {W.~L.}\ \bibnamefont {Yang}}, \bibinfo {author}
  {\bibfnamefont {F.}~\bibnamefont {Ronning}}, \bibinfo {author} {\bibfnamefont
  {T.}~\bibnamefont {Sasagawa}}, \bibinfo {author} {\bibfnamefont
  {T.}~\bibnamefont {Kakeshita}}, \bibinfo {author} {\bibfnamefont
  {T.}~\bibnamefont {Noda}}, \bibinfo {author} {\bibfnamefont {H.}~\bibnamefont
  {Eisaki}}, \bibinfo {author} {\bibfnamefont {S.}~\bibnamefont {Uchida}},
  \bibinfo {author} {\bibfnamefont {C.~T.}\ \bibnamefont {Lin}}, \bibinfo
  {author} {\bibfnamefont {F.}~\bibnamefont {Zhou}}, \bibinfo {author}
  {\bibfnamefont {J.~W.}\ \bibnamefont {Xiong}}, \bibinfo {author}
  {\bibfnamefont {W.~X.}\ \bibnamefont {Ti}}, \bibinfo {author} {\bibfnamefont
  {Z.~X.}\ \bibnamefont {Zhao}}, \bibinfo {author} {\bibfnamefont
  {A.}~\bibnamefont {Fujimori}}, \bibinfo {author} {\bibfnamefont
  {Z.}~\bibnamefont {Hussain}}, \ and\ \bibinfo {author} {\bibfnamefont
  {Z.-X.}\ \bibnamefont {Shen}},\ }\href {\doibase 10.1038/423398a} {\bibfield
  {journal} {\bibinfo  {journal} {Nature}\ }\textbf {\bibinfo {volume} {423}},\
  \bibinfo {pages} {398} (\bibinfo {year} {2003})}\BibitemShut {NoStop}%
\bibitem [{\citenamefont {Vishik}\ \emph
  {et~al.}(2010{\natexlab{a}})\citenamefont {Vishik}, \citenamefont {Lee},
  \citenamefont {Schmitt}, \citenamefont {Moritz}, \citenamefont {Sasagawa},
  \citenamefont {Uchida}, \citenamefont {Fujita}, \citenamefont {Ishida},
  \citenamefont {Zhang}, \citenamefont {Devereaux},\ and\ \citenamefont
  {Shen}}]{Vishik2}%
  \BibitemOpen
  \bibfield  {author} {\bibinfo {author} {\bibfnamefont {I.~M.}\ \bibnamefont
  {Vishik}}, \bibinfo {author} {\bibfnamefont {W.~S.}\ \bibnamefont {Lee}},
  \bibinfo {author} {\bibfnamefont {F.}~\bibnamefont {Schmitt}}, \bibinfo
  {author} {\bibfnamefont {B.}~\bibnamefont {Moritz}}, \bibinfo {author}
  {\bibfnamefont {T.}~\bibnamefont {Sasagawa}}, \bibinfo {author}
  {\bibfnamefont {S.}~\bibnamefont {Uchida}}, \bibinfo {author} {\bibfnamefont
  {K.}~\bibnamefont {Fujita}}, \bibinfo {author} {\bibfnamefont
  {S.}~\bibnamefont {Ishida}}, \bibinfo {author} {\bibfnamefont
  {C.}~\bibnamefont {Zhang}}, \bibinfo {author} {\bibfnamefont {T.~P.}\
  \bibnamefont {Devereaux}}, \ and\ \bibinfo {author} {\bibfnamefont {Z.~X.}\
  \bibnamefont {Shen}},\ }\href {\doibase 10.1103/PhysRevLett.104.207002}
  {\bibfield  {journal} {\bibinfo  {journal} {Phys. Rev. Lett.}\ }\textbf
  {\bibinfo {volume} {104}},\ \bibinfo {pages} {207002} (\bibinfo {year}
  {2010}{\natexlab{a}})}\BibitemShut {NoStop}%
\bibitem [{\citenamefont {Johnston}\ \emph {et~al.}(2012)\citenamefont
  {Johnston}, \citenamefont {Vishik}, \citenamefont {Lee}, \citenamefont
  {Schmitt}, \citenamefont {Uchida}, \citenamefont {Fujita}, \citenamefont
  {Ishida}, \citenamefont {Nagaosa}, \citenamefont {Shen},\ and\ \citenamefont
  {Devereaux}}]{Johnston}%
  \BibitemOpen
  \bibfield  {author} {\bibinfo {author} {\bibfnamefont {S.}~\bibnamefont
  {Johnston}}, \bibinfo {author} {\bibfnamefont {I.~M.}\ \bibnamefont
  {Vishik}}, \bibinfo {author} {\bibfnamefont {W.~S.}\ \bibnamefont {Lee}},
  \bibinfo {author} {\bibfnamefont {F.}~\bibnamefont {Schmitt}}, \bibinfo
  {author} {\bibfnamefont {S.}~\bibnamefont {Uchida}}, \bibinfo {author}
  {\bibfnamefont {K.}~\bibnamefont {Fujita}}, \bibinfo {author} {\bibfnamefont
  {S.}~\bibnamefont {Ishida}}, \bibinfo {author} {\bibfnamefont
  {N.}~\bibnamefont {Nagaosa}}, \bibinfo {author} {\bibfnamefont {Z.~X.}\
  \bibnamefont {Shen}}, \ and\ \bibinfo {author} {\bibfnamefont {T.~P.}\
  \bibnamefont {Devereaux}},\ }\href {\doibase 10.1103/PhysRevLett.108.166404}
  {\bibfield  {journal} {\bibinfo  {journal} {Phys. Rev. Lett.}\ }\textbf
  {\bibinfo {volume} {108}},\ \bibinfo {pages} {166404} (\bibinfo {year}
  {2012})}\BibitemShut {NoStop}%
\bibitem [{\citenamefont {Paramekanti}\ \emph {et~al.}(2001)\citenamefont
  {Paramekanti}, \citenamefont {Randeria},\ and\ \citenamefont
  {Trivedi}}]{Paramekanti}%
  \BibitemOpen
  \bibfield  {author} {\bibinfo {author} {\bibfnamefont {A.}~\bibnamefont
  {Paramekanti}}, \bibinfo {author} {\bibfnamefont {M.}~\bibnamefont
  {Randeria}}, \ and\ \bibinfo {author} {\bibfnamefont {N.}~\bibnamefont
  {Trivedi}},\ }\href {\doibase 10.1103/PhysRevLett.87.217002} {\bibfield
  {journal} {\bibinfo  {journal} {Phys. Rev. Lett.}\ }\textbf {\bibinfo
  {volume} {87}},\ \bibinfo {pages} {217002} (\bibinfo {year}
  {2001})}\BibitemShut {NoStop}%
\bibitem [{\citenamefont {Randeria}\ \emph {et~al.}(2004)\citenamefont
  {Randeria}, \citenamefont {Paramekanti},\ and\ \citenamefont
  {Trivedi}}]{PhysRevB.69.144509}%
  \BibitemOpen
  \bibfield  {author} {\bibinfo {author} {\bibfnamefont {M.}~\bibnamefont
  {Randeria}}, \bibinfo {author} {\bibfnamefont {A.}~\bibnamefont
  {Paramekanti}}, \ and\ \bibinfo {author} {\bibfnamefont {N.}~\bibnamefont
  {Trivedi}},\ }\href {\doibase 10.1103/PhysRevB.69.144509} {\bibfield
  {journal} {\bibinfo  {journal} {Phys. Rev. B}\ }\textbf {\bibinfo {volume}
  {69}},\ \bibinfo {pages} {144509} (\bibinfo {year} {2004})}\BibitemShut
  {NoStop}%
\bibitem [{\citenamefont {Paramekanti}\ \emph {et~al.}(2004)\citenamefont
  {Paramekanti}, \citenamefont {Randeria},\ and\ \citenamefont
  {Trivedi}}]{PhysRevB.70.054504}%
  \BibitemOpen
  \bibfield  {author} {\bibinfo {author} {\bibfnamefont {A.}~\bibnamefont
  {Paramekanti}}, \bibinfo {author} {\bibfnamefont {M.}~\bibnamefont
  {Randeria}}, \ and\ \bibinfo {author} {\bibfnamefont {N.}~\bibnamefont
  {Trivedi}},\ }\href {\doibase 10.1103/PhysRevB.70.054504} {\bibfield
  {journal} {\bibinfo  {journal} {Phys. Rev. B}\ }\textbf {\bibinfo {volume}
  {70}},\ \bibinfo {pages} {054504} (\bibinfo {year} {2004})}\BibitemShut
  {NoStop}%
\bibitem [{\citenamefont {Campuzano}\ \emph {et~al.}(1999)\citenamefont
  {Campuzano}, \citenamefont {Ding}, \citenamefont {Norman}, \citenamefont
  {Fretwell}, \citenamefont {Randeria}, \citenamefont {Kaminski}, \citenamefont
  {Mesot}, \citenamefont {Takeuchi}, \citenamefont {Sato}, \citenamefont
  {Yokoya}, \citenamefont {Takahashi}, \citenamefont {Mochiku}, \citenamefont
  {Kadowaki}, \citenamefont {Guptasarma}, \citenamefont {Hinks}, \citenamefont
  {Konstantinovic}, \citenamefont {Li},\ and\ \citenamefont
  {Raffy}}]{PhysRevLett.83.3709}%
  \BibitemOpen
  \bibfield  {author} {\bibinfo {author} {\bibfnamefont {J.~C.}\ \bibnamefont
  {Campuzano}}, \bibinfo {author} {\bibfnamefont {H.}~\bibnamefont {Ding}},
  \bibinfo {author} {\bibfnamefont {M.~R.}\ \bibnamefont {Norman}}, \bibinfo
  {author} {\bibfnamefont {H.~M.}\ \bibnamefont {Fretwell}}, \bibinfo {author}
  {\bibfnamefont {M.}~\bibnamefont {Randeria}}, \bibinfo {author}
  {\bibfnamefont {A.}~\bibnamefont {Kaminski}}, \bibinfo {author}
  {\bibfnamefont {J.}~\bibnamefont {Mesot}}, \bibinfo {author} {\bibfnamefont
  {T.}~\bibnamefont {Takeuchi}}, \bibinfo {author} {\bibfnamefont
  {T.}~\bibnamefont {Sato}}, \bibinfo {author} {\bibfnamefont {T.}~\bibnamefont
  {Yokoya}}, \bibinfo {author} {\bibfnamefont {T.}~\bibnamefont {Takahashi}},
  \bibinfo {author} {\bibfnamefont {T.}~\bibnamefont {Mochiku}}, \bibinfo
  {author} {\bibfnamefont {K.}~\bibnamefont {Kadowaki}}, \bibinfo {author}
  {\bibfnamefont {P.}~\bibnamefont {Guptasarma}}, \bibinfo {author}
  {\bibfnamefont {D.~G.}\ \bibnamefont {Hinks}}, \bibinfo {author}
  {\bibfnamefont {Z.}~\bibnamefont {Konstantinovic}}, \bibinfo {author}
  {\bibfnamefont {Z.~Z.}\ \bibnamefont {Li}}, \ and\ \bibinfo {author}
  {\bibfnamefont {H.}~\bibnamefont {Raffy}},\ }\href {\doibase
  10.1103/PhysRevLett.83.3709} {\bibfield  {journal} {\bibinfo  {journal}
  {Phys. Rev. Lett.}\ }\textbf {\bibinfo {volume} {83}},\ \bibinfo {pages}
  {3709} (\bibinfo {year} {1999})}\BibitemShut {NoStop}%
\bibitem [{\citenamefont {Sacuto}\ \emph {et~al.}(2013)\citenamefont {Sacuto},
  \citenamefont {Benhabib}, \citenamefont {Gallais}, \citenamefont {Blanc},
  \citenamefont {Cazayous}, \citenamefont {Méasson}, \citenamefont {Wen},
  \citenamefont {Xu},\ and\ \citenamefont {Gu}}]{Sacuto}%
  \BibitemOpen
  \bibfield  {author} {\bibinfo {author} {\bibfnamefont {A.}~\bibnamefont
  {Sacuto}}, \bibinfo {author} {\bibfnamefont {S.}~\bibnamefont {Benhabib}},
  \bibinfo {author} {\bibfnamefont {Y.}~\bibnamefont {Gallais}}, \bibinfo
  {author} {\bibfnamefont {S.}~\bibnamefont {Blanc}}, \bibinfo {author}
  {\bibfnamefont {M.}~\bibnamefont {Cazayous}}, \bibinfo {author}
  {\bibfnamefont {M.-A.}\ \bibnamefont {Méasson}}, \bibinfo {author}
  {\bibfnamefont {J.~S.}\ \bibnamefont {Wen}}, \bibinfo {author} {\bibfnamefont
  {Z.~J.}\ \bibnamefont {Xu}}, \ and\ \bibinfo {author} {\bibfnamefont {G.~D.}\
  \bibnamefont {Gu}},\ }\href {\doibase 10.1088/1742-6596/449/1/012011}
  {\bibfield  {journal} {\bibinfo  {journal} {J. Phys.: Conf. Ser.}\ }\textbf
  {\bibinfo {volume} {449}},\ \bibinfo {pages} {012011} (\bibinfo {year}
  {2013})}\BibitemShut {NoStop}%
\bibitem [{\citenamefont {Shekhter}\ \emph {et~al.}(2013)\citenamefont
  {Shekhter}, \citenamefont {Ramshaw}, \citenamefont {Liang}, \citenamefont
  {Hardy}, \citenamefont {Bonn}, \citenamefont {Balakirev}, \citenamefont
  {McDonald}, \citenamefont {Betts}, \citenamefont {Riggs},\ and\ \citenamefont
  {Migliori}}]{Shekhter}%
  \BibitemOpen
  \bibfield  {author} {\bibinfo {author} {\bibfnamefont {A.}~\bibnamefont
  {Shekhter}}, \bibinfo {author} {\bibfnamefont {B.~J.}\ \bibnamefont
  {Ramshaw}}, \bibinfo {author} {\bibfnamefont {R.}~\bibnamefont {Liang}},
  \bibinfo {author} {\bibfnamefont {W.~N.}\ \bibnamefont {Hardy}}, \bibinfo
  {author} {\bibfnamefont {D.~A.}\ \bibnamefont {Bonn}}, \bibinfo {author}
  {\bibfnamefont {F.~F.}\ \bibnamefont {Balakirev}}, \bibinfo {author}
  {\bibfnamefont {R.~D.}\ \bibnamefont {McDonald}}, \bibinfo {author}
  {\bibfnamefont {J.~B.}\ \bibnamefont {Betts}}, \bibinfo {author}
  {\bibfnamefont {S.~C.}\ \bibnamefont {Riggs}}, \ and\ \bibinfo {author}
  {\bibfnamefont {A.}~\bibnamefont {Migliori}},\ }\href {\doibase
  10.1038/nature12165} {\bibfield  {journal} {\bibinfo  {journal} {Nature}\
  }\textbf {\bibinfo {volume} {498}},\ \bibinfo {pages} {75} (\bibinfo {year}
  {2013})}\BibitemShut {NoStop}%
\bibitem [{\citenamefont {Efetov}\ \emph {et~al.}(2013)\citenamefont {Efetov},
  \citenamefont {Meier},\ and\ \citenamefont {Pepin}}]{Efetov}%
  \BibitemOpen
  \bibfield  {author} {\bibinfo {author} {\bibfnamefont {K.~B.}\ \bibnamefont
  {Efetov}}, \bibinfo {author} {\bibfnamefont {H.}~\bibnamefont {Meier}}, \
  and\ \bibinfo {author} {\bibfnamefont {C.}~\bibnamefont {Pepin}},\ }\href
  {\doibase 10.1038/nphys2641} {\bibfield  {journal} {\bibinfo  {journal}
  {Nature Phys.}\ }\textbf {\bibinfo {volume} {9}},\ \bibinfo {pages} {442}
  (\bibinfo {year} {2013})}\BibitemShut {NoStop}%
\bibitem [{\citenamefont {Kondo}\ \emph {et~al.}(2009)\citenamefont {Kondo},
  \citenamefont {Khasanov}, \citenamefont {Takeuchi}, \citenamefont
  {Schmalian},\ and\ \citenamefont {Kaminski}}]{KondoT}%
  \BibitemOpen
  \bibfield  {author} {\bibinfo {author} {\bibfnamefont {T.}~\bibnamefont
  {Kondo}}, \bibinfo {author} {\bibfnamefont {R.}~\bibnamefont {Khasanov}},
  \bibinfo {author} {\bibfnamefont {T.}~\bibnamefont {Takeuchi}}, \bibinfo
  {author} {\bibfnamefont {J.}~\bibnamefont {Schmalian}}, \ and\ \bibinfo
  {author} {\bibfnamefont {A.}~\bibnamefont {Kaminski}},\ }\href
  {http://dx.doi.org/10.1038/nature07644} {\bibfield  {journal} {\bibinfo
  {journal} {Nature}\ }\textbf {\bibinfo {volume} {457}},\ \bibinfo {pages}
  {296} (\bibinfo {year} {2009})}\BibitemShut {NoStop}%
\bibitem [{\citenamefont {Kondo}\ \emph {et~al.}(2010)\citenamefont {Kondo},
  \citenamefont {Hamaya}, \citenamefont {Palczewski}, \citenamefont {Takeuchi},
  \citenamefont {Wen}, \citenamefont {Gu}, \citenamefont {Schmalian},\ and\
  \citenamefont {Kaminski}}]{KondoT2}%
  \BibitemOpen
  \bibfield  {author} {\bibinfo {author} {\bibfnamefont {T.}~\bibnamefont
  {Kondo}}, \bibinfo {author} {\bibfnamefont {Y.}~\bibnamefont {Hamaya}},
  \bibinfo {author} {\bibfnamefont {A.~D.}\ \bibnamefont {Palczewski}},
  \bibinfo {author} {\bibfnamefont {T.}~\bibnamefont {Takeuchi}}, \bibinfo
  {author} {\bibfnamefont {J.~S.}\ \bibnamefont {Wen}}, \bibinfo {author}
  {\bibfnamefont {G.}~\bibnamefont {Gu}}, \bibinfo {author} {\bibfnamefont
  {J.}~\bibnamefont {Schmalian}}, \ and\ \bibinfo {author} {\bibfnamefont
  {A.}~\bibnamefont {Kaminski}},\ }\href {\doibase 10.1038/nphys1851}
  {\bibfield  {journal} {\bibinfo  {journal} {Nature Phys.}\ }\textbf {\bibinfo
  {volume} {7}},\ \bibinfo {pages} {21} (\bibinfo {year} {2010})}\BibitemShut
  {NoStop}%
\bibitem [{\citenamefont {Khasanov}\ \emph {et~al.}(2008)\citenamefont
  {Khasanov}, \citenamefont {Kondo}, \citenamefont {Str\"assle}, \citenamefont
  {Heron}, \citenamefont {Kaminski}, \citenamefont {Keller}, \citenamefont
  {Lee},\ and\ \citenamefont {Takeuchi}}]{PhysRevLett.101.227002}%
  \BibitemOpen
  \bibfield  {author} {\bibinfo {author} {\bibfnamefont {R.}~\bibnamefont
  {Khasanov}}, \bibinfo {author} {\bibfnamefont {T.}~\bibnamefont {Kondo}},
  \bibinfo {author} {\bibfnamefont {S.}~\bibnamefont {Str\"assle}}, \bibinfo
  {author} {\bibfnamefont {D.~O.~G.}\ \bibnamefont {Heron}}, \bibinfo {author}
  {\bibfnamefont {A.}~\bibnamefont {Kaminski}}, \bibinfo {author}
  {\bibfnamefont {H.}~\bibnamefont {Keller}}, \bibinfo {author} {\bibfnamefont
  {S.~L.}\ \bibnamefont {Lee}}, \ and\ \bibinfo {author} {\bibfnamefont
  {T.}~\bibnamefont {Takeuchi}},\ }\href {\doibase
  10.1103/PhysRevLett.101.227002} {\bibfield  {journal} {\bibinfo  {journal}
  {Phys. Rev. Lett.}\ }\textbf {\bibinfo {volume} {101}},\ \bibinfo {pages}
  {227002} (\bibinfo {year} {2008})}\BibitemShut {NoStop}%
\bibitem [{\citenamefont {Yokoyama}\ \emph {et~al.}(2013)\citenamefont
  {Yokoyama}, \citenamefont {Ogata}, \citenamefont {Tanaka}, \citenamefont
  {Kobayashi},\ and\ \citenamefont {Tsuchiura}}]{Yokoyama2}%
  \BibitemOpen
  \bibfield  {author} {\bibinfo {author} {\bibfnamefont {H.}~\bibnamefont
  {Yokoyama}}, \bibinfo {author} {\bibfnamefont {M.}~\bibnamefont {Ogata}},
  \bibinfo {author} {\bibfnamefont {Y.}~\bibnamefont {Tanaka}}, \bibinfo
  {author} {\bibfnamefont {K.}~\bibnamefont {Kobayashi}}, \ and\ \bibinfo
  {author} {\bibfnamefont {H.}~\bibnamefont {Tsuchiura}},\ }\href {\doibase
  10.1143/JPSJ.82.014707} {\bibfield  {journal} {\bibinfo  {journal} {J. Phys.
  Soc. Jpn.}\ }\textbf {\bibinfo {volume} {82}},\ \bibinfo {pages} {014707}
  (\bibinfo {year} {2013})}\BibitemShut {NoStop}%
\bibitem [{\citenamefont {Yamase}\ and\ \citenamefont
  {Kohno}(2000{\natexlab{a}})}]{JPSJ.69.2151}%
  \BibitemOpen
  \bibfield  {author} {\bibinfo {author} {\bibfnamefont {H.}~\bibnamefont
  {Yamase}}\ and\ \bibinfo {author} {\bibfnamefont {H.}~\bibnamefont {Kohno}},\
  }\href {\doibase 10.1143/JPSJ.69.2151} {\bibfield  {journal} {\bibinfo
  {journal} {Journal of the Physical Society of Japan}\ }\textbf {\bibinfo
  {volume} {69}},\ \bibinfo {pages} {2151} (\bibinfo {year}
  {2000}{\natexlab{a}})}\BibitemShut {NoStop}%
\bibitem [{\citenamefont {Yamase}\ and\ \citenamefont
  {Kohno}(2000{\natexlab{b}})}]{JPSJ.69.332}%
  \BibitemOpen
  \bibfield  {author} {\bibinfo {author} {\bibfnamefont {H.}~\bibnamefont
  {Yamase}}\ and\ \bibinfo {author} {\bibfnamefont {H.}~\bibnamefont {Kohno}},\
  }\href {\doibase 10.1143/JPSJ.69.332} {\bibfield  {journal} {\bibinfo
  {journal} {Journal of the Physical Society of Japan}\ }\textbf {\bibinfo
  {volume} {69}},\ \bibinfo {pages} {332} (\bibinfo {year}
  {2000}{\natexlab{b}})}\BibitemShut {NoStop}%
\bibitem [{\citenamefont {Halboth}\ and\ \citenamefont
  {Metzner}(2000)}]{PhysRevLett.85.5162}%
  \BibitemOpen
  \bibfield  {author} {\bibinfo {author} {\bibfnamefont {C.~J.}\ \bibnamefont
  {Halboth}}\ and\ \bibinfo {author} {\bibfnamefont {W.}~\bibnamefont
  {Metzner}},\ }\href {\doibase 10.1103/PhysRevLett.85.5162} {\bibfield
  {journal} {\bibinfo  {journal} {Phys. Rev. Lett.}\ }\textbf {\bibinfo
  {volume} {85}},\ \bibinfo {pages} {5162} (\bibinfo {year}
  {2000})}\BibitemShut {NoStop}%
\bibitem [{\citenamefont {Zheng}\ \emph {et~al.}()\citenamefont {Zheng},
  \citenamefont {Huang},\ and\ \citenamefont {Zou}}]{Zheng}%
  \BibitemOpen
  \bibfield  {author} {\bibinfo {author} {\bibfnamefont {X.-J.}\ \bibnamefont
  {Zheng}}, \bibinfo {author} {\bibfnamefont {Z.-B.}\ \bibnamefont {Huang}}, \
  and\ \bibinfo {author} {\bibfnamefont {L.-J.}\ \bibnamefont {Zou}},\
  }\href@noop {} {}\bibinfo {howpublished} {arXiv:1301.1012}\BibitemShut
  {NoStop}%
\bibitem [{\citenamefont {Mesot}\ \emph {et~al.}(1999)\citenamefont {Mesot},
  \citenamefont {Norman}, \citenamefont {Ding}, \citenamefont {Randeria},
  \citenamefont {Campuzano}, \citenamefont {Paramekanti}, \citenamefont
  {Fretwell}, \citenamefont {Kaminski}, \citenamefont {Takeuchi}, \citenamefont
  {Yokoya}, \citenamefont {Sato}, \citenamefont {Takahashi}, \citenamefont
  {Mochiku},\ and\ \citenamefont {Kadowaki}}]{Mesot}%
  \BibitemOpen
  \bibfield  {author} {\bibinfo {author} {\bibfnamefont {J.}~\bibnamefont
  {Mesot}}, \bibinfo {author} {\bibfnamefont {M.~R.}\ \bibnamefont {Norman}},
  \bibinfo {author} {\bibfnamefont {H.}~\bibnamefont {Ding}}, \bibinfo {author}
  {\bibfnamefont {M.}~\bibnamefont {Randeria}}, \bibinfo {author}
  {\bibfnamefont {J.~C.}\ \bibnamefont {Campuzano}}, \bibinfo {author}
  {\bibfnamefont {A.}~\bibnamefont {Paramekanti}}, \bibinfo {author}
  {\bibfnamefont {H.~M.}\ \bibnamefont {Fretwell}}, \bibinfo {author}
  {\bibfnamefont {A.}~\bibnamefont {Kaminski}}, \bibinfo {author}
  {\bibfnamefont {T.}~\bibnamefont {Takeuchi}}, \bibinfo {author}
  {\bibfnamefont {T.}~\bibnamefont {Yokoya}}, \bibinfo {author} {\bibfnamefont
  {T.}~\bibnamefont {Sato}}, \bibinfo {author} {\bibfnamefont {T.}~\bibnamefont
  {Takahashi}}, \bibinfo {author} {\bibfnamefont {T.}~\bibnamefont {Mochiku}},
  \ and\ \bibinfo {author} {\bibfnamefont {K.}~\bibnamefont {Kadowaki}},\
  }\href {\doibase 10.1103/PhysRevLett.83.840} {\bibfield  {journal} {\bibinfo
  {journal} {Phys. Rev. Lett.}\ }\textbf {\bibinfo {volume} {83}},\ \bibinfo
  {pages} {840} (\bibinfo {year} {1999})}\BibitemShut {NoStop}%
\bibitem [{\citenamefont {Yoshida}\ \emph {et~al.}(2012)\citenamefont
  {Yoshida}, \citenamefont {Hashimoto}, \citenamefont {Vishik}, \citenamefont
  {Shen},\ and\ \citenamefont {Fujimori}}]{Yoshida}%
  \BibitemOpen
  \bibfield  {author} {\bibinfo {author} {\bibfnamefont {T.}~\bibnamefont
  {Yoshida}}, \bibinfo {author} {\bibfnamefont {M.}~\bibnamefont {Hashimoto}},
  \bibinfo {author} {\bibfnamefont {I.~M.}\ \bibnamefont {Vishik}}, \bibinfo
  {author} {\bibfnamefont {Z.-X.}\ \bibnamefont {Shen}}, \ and\ \bibinfo
  {author} {\bibfnamefont {A.}~\bibnamefont {Fujimori}},\ }\href {\doibase
  10.1143/JPSJ.81.011006} {\bibfield  {journal} {\bibinfo  {journal} {J. Phys.
  Soc. Jpn.}\ }\textbf {\bibinfo {volume} {81}},\ \bibinfo {pages} {011006}
  (\bibinfo {year} {2012})}\BibitemShut {NoStop}%
\bibitem [{\citenamefont {Lee}\ \emph {et~al.}(2007)\citenamefont {Lee},
  \citenamefont {Vishik}, \citenamefont {Tanaka}, \citenamefont {Lu},
  \citenamefont {Sasagawa}, \citenamefont {Nagaosa}, \citenamefont {Devereaux},
  \citenamefont {Hussain},\ and\ \citenamefont {Shen}}]{Lee}%
  \BibitemOpen
  \bibfield  {author} {\bibinfo {author} {\bibfnamefont {W.}~\bibnamefont
  {Lee}}, \bibinfo {author} {\bibfnamefont {I.}~\bibnamefont {Vishik}},
  \bibinfo {author} {\bibfnamefont {K.}~\bibnamefont {Tanaka}}, \bibinfo
  {author} {\bibfnamefont {D.}~\bibnamefont {Lu}}, \bibinfo {author}
  {\bibfnamefont {T.}~\bibnamefont {Sasagawa}}, \bibinfo {author}
  {\bibfnamefont {N.}~\bibnamefont {Nagaosa}}, \bibinfo {author} {\bibfnamefont
  {T.}~\bibnamefont {Devereaux}}, \bibinfo {author} {\bibfnamefont
  {Z.}~\bibnamefont {Hussain}}, \ and\ \bibinfo {author} {\bibfnamefont
  {Z.-X.}\ \bibnamefont {Shen}},\ }\href@noop {} {\bibfield  {journal}
  {\bibinfo  {journal} {Nature}\ }\textbf {\bibinfo {volume} {450}},\ \bibinfo
  {pages} {81} (\bibinfo {year} {2007})}\BibitemShut {NoStop}%
\bibitem [{\citenamefont {Vishik}\ \emph
  {et~al.}(2010{\natexlab{b}})\citenamefont {Vishik}, \citenamefont {Lee},
  \citenamefont {He}, \citenamefont {Hashimoto}, \citenamefont {Hussain},
  \citenamefont {Devereaux},\ and\ \citenamefont {Shen}}]{Vishik}%
  \BibitemOpen
  \bibfield  {author} {\bibinfo {author} {\bibfnamefont {I.~M.}\ \bibnamefont
  {Vishik}}, \bibinfo {author} {\bibfnamefont {W.~S.}\ \bibnamefont {Lee}},
  \bibinfo {author} {\bibfnamefont {R.-H.}\ \bibnamefont {He}}, \bibinfo
  {author} {\bibfnamefont {M.}~\bibnamefont {Hashimoto}}, \bibinfo {author}
  {\bibfnamefont {Z.}~\bibnamefont {Hussain}}, \bibinfo {author} {\bibfnamefont
  {T.~P.}\ \bibnamefont {Devereaux}}, \ and\ \bibinfo {author} {\bibfnamefont
  {Z.-X.}\ \bibnamefont {Shen}},\ }\href@noop {} {\bibfield  {journal}
  {\bibinfo  {journal} {New J. Phys.}\ }\textbf {\bibinfo {volume} {12}},\
  \bibinfo {pages} {105008} (\bibinfo {year} {2010}{\natexlab{b}})}\BibitemShut
  {NoStop}%
\bibitem [{\citenamefont {Vishik}\ \emph {et~al.}(2012)\citenamefont {Vishik},
  \citenamefont {Hashimoto}, \citenamefont {He}, \citenamefont {Lee},
  \citenamefont {Schmitt}, \citenamefont {Lu}, \citenamefont {Moore},
  \citenamefont {Zhang}, \citenamefont {Meevasana}, \citenamefont {Sasagawa},
  \citenamefont {Uchida}, \citenamefont {Fujita}, \citenamefont {Ishida},
  \citenamefont {Ishikado}, \citenamefont {Yoshida}, \citenamefont {Eisaki},
  \citenamefont {Hussain}, \citenamefont {Devereaux},\ and\ \citenamefont
  {Shen}}]{Vishik3}%
  \BibitemOpen
  \bibfield  {author} {\bibinfo {author} {\bibfnamefont {I.~M.}\ \bibnamefont
  {Vishik}}, \bibinfo {author} {\bibfnamefont {M.}~\bibnamefont {Hashimoto}},
  \bibinfo {author} {\bibfnamefont {R.-H.}\ \bibnamefont {He}}, \bibinfo
  {author} {\bibfnamefont {W.-S.}\ \bibnamefont {Lee}}, \bibinfo {author}
  {\bibfnamefont {F.}~\bibnamefont {Schmitt}}, \bibinfo {author} {\bibfnamefont
  {D.}~\bibnamefont {Lu}}, \bibinfo {author} {\bibfnamefont {R.~G.}\
  \bibnamefont {Moore}}, \bibinfo {author} {\bibfnamefont {C.}~\bibnamefont
  {Zhang}}, \bibinfo {author} {\bibfnamefont {W.}~\bibnamefont {Meevasana}},
  \bibinfo {author} {\bibfnamefont {T.}~\bibnamefont {Sasagawa}}, \bibinfo
  {author} {\bibfnamefont {S.}~\bibnamefont {Uchida}}, \bibinfo {author}
  {\bibfnamefont {K.}~\bibnamefont {Fujita}}, \bibinfo {author} {\bibfnamefont
  {S.}~\bibnamefont {Ishida}}, \bibinfo {author} {\bibfnamefont
  {M.}~\bibnamefont {Ishikado}}, \bibinfo {author} {\bibfnamefont
  {Y.}~\bibnamefont {Yoshida}}, \bibinfo {author} {\bibfnamefont
  {H.}~\bibnamefont {Eisaki}}, \bibinfo {author} {\bibfnamefont
  {Z.}~\bibnamefont {Hussain}}, \bibinfo {author} {\bibfnamefont {T.~P.}\
  \bibnamefont {Devereaux}}, \ and\ \bibinfo {author} {\bibfnamefont {Z.-X.}\
  \bibnamefont {Shen}},\ }\href {\doibase 10.1073/pnas.1209471109} {\bibfield
  {journal} {\bibinfo  {journal} {Proc. Natl. Acad. Sci. USA}\ }\textbf
  {\bibinfo {volume} {109}},\ \bibinfo {pages} {18332} (\bibinfo {year}
  {2012})}\BibitemShut {NoStop}%
\bibitem [{\citenamefont {Feng}\ \emph {et~al.}(2012)\citenamefont {Feng},
  \citenamefont {Zhao},\ and\ \citenamefont {Huang}}]{PhysRevB.85.054509}%
  \BibitemOpen
  \bibfield  {author} {\bibinfo {author} {\bibfnamefont {S.}~\bibnamefont
  {Feng}}, \bibinfo {author} {\bibfnamefont {H.}~\bibnamefont {Zhao}}, \ and\
  \bibinfo {author} {\bibfnamefont {Z.}~\bibnamefont {Huang}},\ }\href
  {\doibase 10.1103/PhysRevB.85.054509} {\bibfield  {journal} {\bibinfo
  {journal} {Phys. Rev. B}\ }\textbf {\bibinfo {volume} {85}},\ \bibinfo
  {pages} {054509} (\bibinfo {year} {2012})}\BibitemShut {NoStop}%
\bibitem [{\citenamefont {Yokoyama}\ \emph {et~al.}(2004)\citenamefont
  {Yokoyama}, \citenamefont {Tanaka}, \citenamefont {Ogata},\ and\
  \citenamefont {Tsuchiura}}]{Yokoyama1}%
  \BibitemOpen
  \bibfield  {author} {\bibinfo {author} {\bibfnamefont {H.}~\bibnamefont
  {Yokoyama}}, \bibinfo {author} {\bibfnamefont {Y.}~\bibnamefont {Tanaka}},
  \bibinfo {author} {\bibfnamefont {M.}~\bibnamefont {Ogata}}, \ and\ \bibinfo
  {author} {\bibfnamefont {H.}~\bibnamefont {Tsuchiura}},\ }\href {\doibase
  10.1143/JPSJ.73.1119} {\bibfield  {journal} {\bibinfo  {journal} {J. Phys.
  Soc. Jpn.}\ }\textbf {\bibinfo {volume} {73}},\ \bibinfo {pages} {1119}
  (\bibinfo {year} {2004})}\BibitemShut {NoStop}%
\bibitem [{\citenamefont {Baeriswyl}(2000)}]{Baeriswyl}%
  \BibitemOpen
  \bibfield  {author} {\bibinfo {author} {\bibfnamefont {D.}~\bibnamefont
  {Baeriswyl}},\ }\href {\doibase 10.1023/A:1003785323041} {\bibfield
  {journal} {\bibinfo  {journal} {Found. Phys.}\ }\textbf {\bibinfo {volume}
  {30}},\ \bibinfo {pages} {2033} (\bibinfo {year} {2000})}\BibitemShut
  {NoStop}%
\bibitem [{\citenamefont {Het\'enyi}(2010)}]{Hetenyi}%
  \BibitemOpen
  \bibfield  {author} {\bibinfo {author} {\bibfnamefont {B.}~\bibnamefont
  {Het\'enyi}},\ }\href {\doibase 10.1103/PhysRevB.82.115104} {\bibfield
  {journal} {\bibinfo  {journal} {Phys. Rev. B}\ }\textbf {\bibinfo {volume}
  {82}},\ \bibinfo {pages} {115104} (\bibinfo {year} {2010})}\BibitemShut
  {NoStop}%
\bibitem [{\citenamefont {Eichenberger}\ and\ \citenamefont
  {Baeriswyl}(2007)}]{Eichenberger}%
  \BibitemOpen
  \bibfield  {author} {\bibinfo {author} {\bibfnamefont {D.}~\bibnamefont
  {Eichenberger}}\ and\ \bibinfo {author} {\bibfnamefont {D.}~\bibnamefont
  {Baeriswyl}},\ }\href {\doibase 10.1103/PhysRevB.76.180504} {\bibfield
  {journal} {\bibinfo  {journal} {Phys. Rev. B}\ }\textbf {\bibinfo {volume}
  {76}},\ \bibinfo {pages} {180504} (\bibinfo {year} {2007})}\BibitemShut
  {NoStop}%
\bibitem [{\citenamefont {Civelli}\ \emph {et~al.}(2008)\citenamefont
  {Civelli}, \citenamefont {Capone}, \citenamefont {Georges}, \citenamefont
  {Haule}, \citenamefont {Parcollet}, \citenamefont {Stanescu},\ and\
  \citenamefont {Kotliar}}]{Civelli1}%
  \BibitemOpen
  \bibfield  {author} {\bibinfo {author} {\bibfnamefont {M.}~\bibnamefont
  {Civelli}}, \bibinfo {author} {\bibfnamefont {M.}~\bibnamefont {Capone}},
  \bibinfo {author} {\bibfnamefont {A.}~\bibnamefont {Georges}}, \bibinfo
  {author} {\bibfnamefont {K.}~\bibnamefont {Haule}}, \bibinfo {author}
  {\bibfnamefont {O.}~\bibnamefont {Parcollet}}, \bibinfo {author}
  {\bibfnamefont {T.~D.}\ \bibnamefont {Stanescu}}, \ and\ \bibinfo {author}
  {\bibfnamefont {G.}~\bibnamefont {Kotliar}},\ }\href {\doibase
  10.1103/PhysRevLett.100.046402} {\bibfield  {journal} {\bibinfo  {journal}
  {Phys. Rev. Lett.}\ }\textbf {\bibinfo {volume} {100}},\ \bibinfo {pages}
  {046402} (\bibinfo {year} {2008})}\BibitemShut {NoStop}%
\bibitem [{\citenamefont {Spa{\l}ek}()}]{JSunpublished}%
  \BibitemOpen
  \bibfield  {author} {\bibinfo {author} {\bibfnamefont {J.}~\bibnamefont
  {Spa{\l}ek}},\ }\href@noop {} {\bibinfo  {journal} {unpublished}\
  }\BibitemShut {NoStop}%
\end{thebibliography}%

\end{document}